\documentclass[10.5pt,amsmath,amssymb,superscriptaddress,floatfix,nofootinbib]{revtex4}
\usepackage[utf8]{inputenc}
\usepackage{epsfig}
\usepackage{amsmath, empheq, amsfonts, amssymb}
\usepackage{graphicx}
\usepackage{hyperref}
\usepackage{tabularx}
\usepackage{float}

\begin{document}

\title{A non-relativistic model for the $[cc][\bar{c}\bar{c}]$ tetraquark}

\author{V.~R.~Debastiani}
\email{vinicius.rodrigues@ific.uv.es}
\affiliation{Departamento de F\'{\i}sica Te\'orica and IFIC, Centro Mixto
Universidad de
Valencia-CSIC Institutos de Investigaci\'on de Paterna, Aptdo. 22085, 46071
Valencia,
Spain}
\affiliation{Instituto de F\'{\i}sica,
Universidade de S\~{a}o Paulo, C.P. 66318, 05389-970 S\~{a}o Paulo, SP, Brazil}

\author{F.~S.~Navarra}
\email{navarra@if.usp.br}
\affiliation{Instituto de F\'{\i}sica,
Universidade de S\~{a}o Paulo, C.P. 66318, 05389-970 S\~{a}o Paulo, SP, Brazil}
\affiliation{Institut de Physique Th\'eorique, Universit\'e Paris Saclay,\\
CEA, CNRS, F-91191, Gif-sur-Yvette, France}

\date{\today}

\begin{abstract}

We use a non-relativistic model to study the  spectroscopy of a
tetraquark composed of $[cc][\bar{c}\bar{c}]$ in a diquark-antidiquark configuration.
By numerically solving the Schr\"{o}dinger equation with a Cornell-inspired potential, we separate the four-body problem into three two-body problems.
Spin-dependent terms (spin-spin, spin-orbit and tensor) are used to describe the splitting structure of the $c\bar{c}$ spectrum and are also extended to the interaction between diquarks. Recent experimental data on charmonium states are used to fix the parameters of the model and a satisfactory description of the spectrum is obtained.
We find that the spin-dependent interaction is sizable in the diquark-antidiquark system, despite the heavy diquark mass, and also that the diquark has a finite size if treated in the same way as the $c\bar{c}$ systems.
We find that the lowest $S$-wave $T_{4c}$ tetraquarks might be below their thresholds of spontaneous dissociation into low-lying charmonium pairs, while orbital and radial excitations would be mostly above the corresponding charmonium pair thresholds.
Finally,  we repeat the calculations without the confining part of the potential and obtain bound diquarks and bound tetraquarks.
This might be relevant to the study of  exotic charmonium in the quark-gluon plasma.
The $T_{4c}$ states could be investigated in the forthcoming experiments at the LHC and Belle II.\\
\end{abstract}

\maketitle

\section{Introduction}\label{Introduction}

The existence of multiquark states with four or more quarks was proposed
decades ago \cite{tetra1,narrow}.  The early papers on tetraquark configurations
were based on the MIT bag model with  light quarks only. Later on, the tetraquark
picture was extended to heavy quarks \cite{tetra3,tetra4}.
Interest in this subject was renewed in the past decade
due to the experimental observation of
states which are not combinations of three quarks ($qqq$) or of quark and
antiquark ($q \bar{q}$).
These new  states present quantum numbers, masses, decay channels and widths
that cannot be explained with the conventional meson or baryon models (they are therefore called \textit{Exotics})
\cite{newjapa,olsen,espo,Guo,revCharm,revExotic,Kang}.
Some of them were even found to be charged, which establishes unambiguously
their exotic nature \cite{Z4430,revCharged}.

In the present work we focus on tetraquarks composed of a single flavor,
charm quarks only, using a diquark-antidiquark picture $[cc][\bar{c}\bar{c}]$,
which we will call $T_{4c}$ or ``the all-charm tetraquark''.

The first work on the all-charm tetraquark was published in 1975 by
Iwasaki \cite{iwasaki}. In a subsequent paper  Chao studied the $T_{4c}$ in the
diquark-antidiquark picture with orbital excitations, and its production in
$e^+e^-$ annihilation \cite{Chao}, including an interesting analysis of the
possible decay channels. Later, in the eighties and nineties,
several works with different approaches addressed the question of the
existence of this $c \bar{c} c \bar{c}$ state
\cite{Ballot,Lipkin,ht1,brac,sema1}.
In  more recent years, after the discovery of the $X(3872)$, a new series of
theoretical works on the subject appeared
\cite{vary,javier,heupel,Chiu-Lattice,wagner,bicudo,WuT4c,
ChenT4c,Wang:2017jtz,Karliner,Richard:2017vry}.

On the experimental side,
recent  measurements of $J/\psi$ pair production are very
promising  and might be the ideal starting point to search for the
all-charm tetraquark. They
have been studied at the LHC, by the LHCb \cite{Aaij:2011yc,Aaij:2016bqq}, CMS
\cite{Khachatryan:2014iia} and ATLAS \cite{Aaboud:2016fzt} collaborations.
Double  $c\bar{c}$ production has also been observed by the
Belle collaboration \cite{Abe:2002rb}. In particular in Refs. \cite{Aaij:2011yc,Aaij:2016bqq}, one can see that there is an
enhancement in the differential production cross-section for $J/\psi$
pairs between 6 and 8 GeV. Further
investigation of the invariant mass distribution in this energy range with high
statistics would bring very useful information about the possible existence of the
$T_{4c}$.

Most of the predictions for the $T_{4c}$ mass lead to values around 6 GeV,
and therefore  lie well above the experimentally known range for charmonium (which is concentrated within  $3$ - $4.5$ GeV). This energy gap makes the all-charm
tetraquark a special object in the sector of exotic multiquarks.
The most discussed tetraquark
candidates (the $X$, $Y$, $Z$ states) are in  the same mass range as conventional
charmonium states and this can lead to confusion.

The absence of light quarks in the $T_{4c}$  makes it unlikely to be
a meson-meson  molecule, since it is not easy to describe this binding in
terms of pion exchange or light vector meson exchange. If it exists, the
$T_{4c}$ is bound by QCD forces and studying its spectrum will lead to a more
complete understanding of QCD interactions. If it does not exist we have to
understand why.

We will describe the $T_{4c}$ as a two-body non-relativistic
system, made of a $c c$ diquark and a $\bar{c} \bar{c}$ antidiquark,
which interact through a Cornell-like potential. We choose the diquark and
antidiquark to be in the color antitriplet and triplet representations,
respectively.

Why do we choose the Cornell model?
We choose it because it is able to capture
the essential aspects of the heavy quark-antiquark
interactions. It has almost never been too wrong and when it was,
there was something really new happening. Moreover, the quark-antiquark
potential can be continuously improved \cite{willi} and its parameters can be
adjusted so as
to incorporate the most recent experimental information on the charmonium spectrum.
Finally, we will study systems with angular momentum and all kinds of spin
interactions. With more constituents, we may form systems with higher spin and
total angular momentum.  With the Cornell model (unlike in some other approaches)
we can identify the individual contribution of each one of these interactions.

We choose to work with diquarks, not only because they simplify the calculations,
but also because there is some evidence of diquark clustering in baryons. In
the case of heavy diquarks the interaction has a stronger short distance component,
in which the perturbative one-gluon exchange may be attractive. In particular,
the $c c$ diquark became  more interesting after the prediction of the
$T_{cc}$ \cite{tcc} and even more so after the  very recent discovery of
the baryon $\Xi_{cc}^{++}$  \cite{baryoncc}, a $ccu$ state where the charm
diquark may play a role.

In the literature we find some calculations which are very simple and strongly
based on the existing empirical information, as in Ref.~\cite{Karliner}, and
some which are very sophisticated, such as the lattice calculations of Ref.~\cite{wagner} or the QCD sum
rules calculations of Ref.~\cite{ChenT4c}. Our model is at an intermediate level, being more precise than the
estimates made in Ref.~\cite{Karliner} and more transparent than the results found in
Refs.~\cite{wagner,ChenT4c}, where it is very difficult to access the role of spin
interactions. Ideally, all these approaches should converge and the origin of the
remaining discrepancies should be well understood. At the end of this work we
will present a comparison with the results obtained in other approaches.

\begin{figure}[htb!]
\begin{center}
\includegraphics[width=0.40\textwidth]{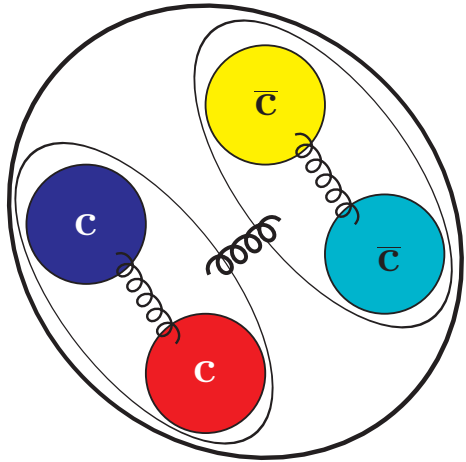}
\caption{\label{fig:T4c}Pictorial representation of the all-charm tetraquark in the
diquark-antiquark scheme.}
\end{center}
\end{figure}

\section{A non-relativistic model}
\label{Formalism}
  \indent

A pictorial representation of the all-charm tetraquark in the
diquark-antiquark scheme of our model can be seen in Figure \ref{fig:T4c}.
One of the most common functional forms of the
zeroth-order potential, $V^{(0)}(r)$, employed in heavy quarkonium
spectroscopy is the \textit{Coulomb plus Linear}, where the Coulomb term
arises from the one-gluon exchange
(OGE) associated with a Lorentz vector structure and the linear part responsible
for confinement, which is usually associated with a Lorentz scalar structure. The potential is
given by:
\begin{equation}
\label{Cornell}
     V_{C+L}^{(0)} = V_{V} + V_{S} \quad \Longrightarrow \quad
    V^{(0)}(r) = \kappa_{s}\frac{\alpha_{s}}{r} + b r
\end{equation}
where $\kappa_{s}$, sometimes called the ``color factor'', is related to the color
configuration of the system (it can be negative or positive), $\alpha_{s}$ is the QCD
fine structure constant and $b$, sometimes called ``string tension'', is related to the strength of the confinement. One could also
add a constant term, which would act as a zero-point energy.

Usually, in heavy quark bound states the kinetic energy of the constituents
is small compared to their rest energy, hence a non-relativistic approach with static potentials can be a reasonable approximation.  In two-body problems involving a central potential, it is convenient to work in the center-of-mass frame (CM),
where one can use spherical coordinates to separate the radial and angular parts of the wavefunction, and the kinetic energy is written in terms of the reduced mass
$\mu = (m_1 m_2)/(m_1+m_2)$.
We start with the time-independent Schr\"{o}dinger equation:
\begin{equation}
    \label{radialHadron}
    \left[\frac{1}{2\mu}\left(-\frac{d^2}{dr^2}
+ \frac{\ell(\ell+1)}{r^2}\right) + V^{(0)}(r)\right]y(r) = E\,y(r).
\end{equation}
We first  solve this $radial$ equation
to obtain the energy eigenstate and the wavefunction of each particular state.
Next, spin-dependent terms are included as perturbative corrections. They account
for
the splitting between states with different quantum numbers. Based on the Breit-Fermi
Hamiltonian for one-gluon exchange \cite{Effec,boundstates,voloshin,beyond}, we
introduce three spin-dependent terms:
$V_{SS}$ (Spin-Spin), $V_{LS}$ (Spin-Orbit) and $V_T$
(Tensor). For equal masses $m_1 = m_2 = m$, they are given by:
\begin{eqnarray}
\label{H:spin-spin}   V_{SS} &=& C_{SS}(r)\; \mathbf{S_1 \cdot S_2} \\ [7pt]
\label{H:spin-orbit}  V_{LS} &=& C_{LS}(r)\; \mathbf{L \cdot S} \\ [7pt]
\label{H:tensor}      V_T  &=& C_{T}(r) \left(  \frac{(\mathbf{S_1 \cdot r})
(\mathbf{S_2 \cdot r})}{\mathbf{r}^2}  - \frac{1}{3}(\mathbf{S_1 \cdot S_2}) \right)
\end{eqnarray}
where the radial-dependent coefficients come from the vector $V_{V}$ and
scalar $V_{S}$
parts of the potential in Eq. (\ref{Cornell}),
\begin{eqnarray}
\label{C:spin-spin}  C_{SS}(r)
  &=&\displaystyle \frac{2}{3m^2} \nabla^2V_V(r)
  = -\frac{8\kappa_s\alpha_s\pi}{3m^2}\delta^3(r) \\ [10pt]
\label{C:spin-orbit}  C_{LS}(r)
  &=&\displaystyle \frac{1}{2m^2}\frac{1}{r} \left[3\frac{dV_V(r)}{dr} -
\frac{dV_S(r)}{dr} \right]
  = -\frac{3\kappa_s\alpha_s}{2m^2}\frac{1}{r^3}
    - \frac{b}{2m^2}\frac{1}{r} \\ [10pt]
\label{C:tensor}  C_{T}(r)
  &=&\displaystyle \frac{1}{m^2} \left[\frac{1}{r}\frac{dV_V(r)}{dr} -
\frac{d^2V_V(r)}{dr^2}\right]
  = -\frac{12\kappa_s\alpha_s}{4m^2} \frac{1}{r^3}
\end{eqnarray}
where $m$ is the constituent mass of the two-body problem (charm quark, or diquark).
The second term  in the spin-orbit correction (proportional to the scalar
contribution) is a Thomas precession,  which follows from the assumption that the
confining interaction comes from a Lorentz scalar structure. Notice that if we introduce a
constant term $V_0$ in the potential, it will not affect these radial coefficients, since
only derivatives appear in them. In fact, adding a constant term only shifts the whole spectrum, forcing a change in the parameters such as to reproduce the charmonium spectrum, without actual improvement in the quality of the fit.

These spin-dependent terms are proportional to $1/m^2$, which justifies their
treatment as first-order perturbation corrections in heavy quark bound states. The expectation
value of their radial-dependent coefficients can be calculated using the
wavefunction obtained with the solution of the Schr\"{o}dinger equation.

This framework appears frequently in quarkonium spectroscopy, but a better agreement
between predicted states and the experimental data for $c\bar{c}$ mesons can be obtained
by
including the spin-spin interaction in the zeroth-order potential used in the
Schr\"{o}dinger equation
(as done in Refs. \cite{GodfreyRelat,Brink-Gauss,T-Barnes,High}),
with the artifact of replacing the Dirac delta 
by a Gaussian function which introduces a new parameter $\sigma$. Then the spin-spin term becomes
\begin{equation}\label{spin2}
V^{(0)}_{SS} = \displaystyle -\frac{8\pi\kappa_s\alpha_s}{3m^2}\left(\frac{\sigma}
{\sqrt{\pi}}\right)^3\mathrm{e}^{-\sigma^2r^2}\mathbf{S_1 \cdot S_2} \\
\end{equation}

When the term $\mathbf{S_1 \cdot S_2}$ acts on the wavefunction it will generate a
constant factor, so we still have a potential as a function  only of the $r$
coordinate.
The expectation value of the operator of the spin-spin interaction can be
calculated in
terms of the spin quantum numbers using the following relation,
$\left< \mathbf{S_1 \cdot S_2} \right> = \left< \frac{1}{2}\left(\mathbf{S}^2 -
\mathbf{S_1}^2 -\mathbf{S_2}^2 \right)\right>$, where $S_1$ and $S_2$ are the spins
of particles 1 and 2 respectively, and $S$ is the total spin in consideration.

The expectation value of the operator of the spin-orbit interaction can be calculated
in terms of the quantum numbers of total angular momentum $J$ (defined by the vector
sum: $\mathbf{J} = \mathbf{L} + \mathbf{S}$), total spin $S$, and orbital angular
momentum $\ell$, using the following relation: $\left< \mathbf{L \cdot S} \right> =
\left< \frac{1}{2}\left(\mathbf{J}^2 -\mathbf{L}^2 -\mathbf{S}^2 \right)\right>$.
For $S$-wave states ($\ell=0$), the spin-orbit term $\langle\mathbf{L \cdot S}\rangle$
is always zero.

The tensor interaction demands a bit of algebra. For convenience, we redefine the
tensor
operator with an extra factor $12$, which we remove from its radial coefficient in Eq.~\eqref{C:tensor}:
\begin{equation} \label{TensorFactor}
\mathbf{S_{12}} \equiv  12\left(  \frac{(\mathbf{S_1 \cdot r})(\mathbf{S_2 \cdot r})}
{\mathbf{r}^2}  - \frac{1}{3}(\mathbf{S_1 \cdot S_2}) \right)
= 4[ 3(\mathbf{S_1 \cdot \hat{r}})(\mathbf{S_2 \cdot \hat{r}})
- \mathbf{S_1 \cdot S_2} ]
\end{equation}
The results for the diagonal matrix elements of the tensor operator between two spin
1/2 particles, like in the $c\bar{c}$ mesons, can be found in Refs. \cite{boundstates,B-S} and also (with more details) in
Ref.~\cite{master}. The expectation value of the tensor is non-zero only for
\begin{align}\label{condTensor}
\begin{aligned}
      &1) \quad \ell\neq0 \quad \mathrm{and} \quad S=1 \quad (\mathrm{triplet}), \\
      &2) \quad J = \ell, \quad \mathrm{or} \quad J = \ell - 1, \quad \mathrm{or}
\quad J = \ell + 1.
\end{aligned}
\end{align}
After some manipulations of the spin operators, with the aid of some relations of
spherical harmonics and the Pauli matrices with respective eigenvalues, we can obtain
the following general result, which satisfies the above conditions
(it always vanishes if $\ell=0$ or $S=0$) for any of the allowed values of $J$ and $\ell$:
\begin{eqnarray}\label{tensorS1/2}
\langle \mathbf{S_{12}} \rangle_{\frac{1}{2}\otimes\frac{1}{2}\rightarrow S=1,\;
\ell\neq0} = \left\{\begin{array}{cl}
\displaystyle -\frac{2\ell}{(2\ell +3)}, &\mbox{if} \quad J = \ell +1,\\
\noalign{\medskip}
\displaystyle +2, &\mbox{if} \quad J = \ell,\\\noalign{\medskip}
\displaystyle -\frac{2(\ell +1)}{(2\ell -1)}, &\mbox{if} \quad J = \ell -1.
\end{array}\right.
\end{eqnarray}
 For instance, for $\ell=1$ we have
$\displaystyle\langle\mathbf{S_{12}}\rangle= -\frac{2}{5},\; +2,\;-4$, for
$J=2,\;1,\;0$, respectively. These results are valid for diagonal matrix
elements. 
 The tensor actually has non-vanishing non-diagonal matrix elements, but as a first-order
perturbation correction they can be neglected. They would be important if the
tensor operator were to be used as part of the potential, which would cause the
mixing of
the wavefunction itself, as in the deuteron \cite{deuteron,Messiah}.

Notice that in order to obtain these three general cases of non-vanishing diagonal
matrix
elements of the tensor operator  for two spin 1/2 particles in Eq.~\eqref{tensorS1/2}, it is necessary to
make use
of a few relations that are valid only for Pauli matrices \cite{B-S,master}, like its eigenvalues and the
anticommutation relation. Therefore, we cannot use this result in the
diquark-antidiquark
tensor interaction (if we wish to treat it as a two-body problem), since the
diquarks can have spin 0 or 1. This issue will be discussed later when we address the tetraquark interaction.

Regarding the wavefunction, we will consider only pure states where
$\ell$ (orbital), $S$ (total spin), and $J$ (total angular momentum) are good quantum
numbers. Then the wavefunction will be composed of a radial part and an angular part
which comes from the coupling of spherical harmonics and spin functions at a specific
value of $J$.

Solving the eigenvalue equation (\ref{radialHadron}), one can obtain the interaction
energy $E$ and the wavefunction $y(r)$ of the two-body system under consideration,
where both depend on the number of nodes of the wavefunction $n$ (or principal quantum
number $N=n+1$), on the orbital angular momentum number $\ell$, and in the case of
the spin-spin correction included in $V^{(0)}$, they will also depend on the total
spin $S$ and on the constituent spins $S_1$ and $S_2$.
Since the Schr\"{o}dinger equation has no analytical solution for the potentials that
are relevant here, we solve it numerically, using an improved version of the code
published in Ref. \cite{lucha}.

An  interesting quantity that can be used to check the validity of the non-relativistic
approximation is the velocity of the constituents in each of the systems in
consideration:  the quark velocity inside the meson or the diquark velocity inside
the tetraquark. As discussed in Ref. \cite{QQpack}, the mean square velocity can be
obtained from the kinetic energy, which can be calculated
directly from the Hamiltonian, or using the Viral Theorem:
\begin{align}
&\langle\mathbf{v}^2\rangle = \frac{1}{2\mu}(E - \langle V^{(0)}(r)\rangle);
&  \langle\mathbf{v}^2\rangle = \frac{1}{4\mu}\langle r \frac{d}{dr} V^{(0)}(r) \rangle
\end{align}
where $V^{(0)}(r)$ is the effective zeroth-order potential placed in the
Schr\"{o}dinger
equation and $\mu$ is the reduced mass:
\begin{equation}\label{mu}
  \mu = \frac{m_1 m_2}{m_1+m_2} = \frac{m}{2}, \quad \mathrm{for} \quad m_1=m_2.
\end{equation}
Both methods yield approximately the same result within the numerical
precision employed.

One interesting aspect of  the non-relativistic approach is that, even though the
charmonium system is not completely non-relativistic, a surprisingly good
reproduction of its mass spectrum can be obtained. As discussed in
Ref. \cite{DeSanctisFIT}, where a charmonium model is developed with completely
relativistic energy and also with non-relativistic kinetic energy,  good agreement
with the experimental data can be obtained with both methods, just by using a
different set of parameters in the effective potential employed.

The value of the square modulus of the wavefunction at the origin, $|\Psi(0)|^2$,
is an important quantity. If the spin-spin interaction was treated as a first-order
perturbation without the Gaussian smearing, it would be proportional to $|\Psi(0)|^2$
because of the Dirac delta. Decay widths can also be calculated using the wavefunction
or its derivative at the origin. Only $S$-wave states ($\ell=0$) have non-zero value of
the wavefunction at the origin. For states with orbital angular momentum
($\ell \neq 0$),
the centrifugal term in the Schr\"{o}dinger equation creates a ``centrifugal barrier'',
which makes the wavefunction at the origin vanish. Thus, for $\ell \neq 0$ we will
assume $|\Psi(0)|^2 = 0$ and for $S$-wave we have
\begin{equation}\label{psi0}
  |\Psi(0)|^2 = |Y^0_0(\theta,\phi)R_{n,\ell}(0)|^2 = \frac{|R_{n,\ell}(0)|^2}{4\pi},
\quad \mathrm{for} \quad \ell = 0.
\end{equation}
In fact, the important quantity  is the square modulus of the \textit{radial}
wavefunction at the origin $|R_{n,\ell}(0)|^2$, which can be obtained directly
from the numerical calculations. In the literature on quarkonium models,
we find the following formula (see Ref.~\cite{boundstates} for a deduction)
that relates the wavefunction at the
origin $|\Psi(0)|^2$ to the radial potential $V^{(0)}(r)$:
\begin{equation}
\label{Psi0}
|\Psi(0)|^2 = \frac{\mu}{2\pi} \Big{\langle} \frac{d}{dr} V^{(0)}(r)\Big{\rangle}
\quad \Longrightarrow \quad |R(0)|^2 = 2\mu \Big{\langle} \frac{d}{dr}V^{(0)}(r)
\Big{\rangle}.
\end{equation}
We have checked that the result obtained directly from the numerical method is
compatible
with the one obtained using the formula above.

In more sophisticated quarkonium models, such as the relativized potential model of
Ref.~\cite{GodfreyRelat}, the coupling constant $\alpha_s$ is considered as
a ``running'' parameter, that changes according to the energy scale of each bound
state. However, we have chosen to adopt the non-relativistic model of Ref. \cite{High},
where $\alpha_s$ is as a constant in the potential, which is also a common approach in
many charmonium models.

The values of $\alpha_s$, 
the charm quark mass $m_c$, the string tension $b$, and the Gaussian parameter
$\sigma$, will be obtained from a fit of the charmonium experimental data,
 and once the best set is found, they are kept fixed to generate the whole mass spectrum.

\subsection{Charmonium}
\indent

In order to get good estimates for  diquarks and tetraquarks, we first study the
spectrum of charmonium. In this case, the color factor $\kappa_s$ in Eq.
(\ref{Cornell}) should be that of a color singlet state, since for
$c\bar{c}$ mesons we have $|q\overline{q}\rangle : 3 \otimes \overline{3} = 1 \oplus 8$ \cite{Grif:Particles,muta}. The result for the color singlet is $\kappa_s= -4/3$ \cite{master,Grif:Particles}.

After having solved the Schr\"{o}dinger equation, the  mass of
a particular state  will be given by:
\begin{equation}
\label{eq.charmonium}
M(c\bar{c}) = 2m_c + E_{c\bar{c}} \;+\; \langle V^{(1)}_{Spin} \rangle_{c\bar{c}}
\end{equation}
The parity  and charge conjugation quantum numbers of $q \bar{q}$ states are given by
\cite{Grif:Particles}  $P = (-1)^{\ell+1}$ and $C= (-1)^{\ell+S}$ repectively.
Using the equation above we calculate the masses $M^{calc}_i$ of the $i$ states with
well defined $P$ and $C$, then we fit the experimentally measured masses $M_i^{exp}$  and
determine the parameters minimizing the $\chi^2$, defined as:
\begin{equation}\label{chi2}
\chi^2 = \sum_i^n (M_i^{calc} - M_i^{exp})^2 \cdot w_i
\end{equation}
Following Refs. \cite{DeSanctisFIT, High}  we choose $w_i=1$, which is equivalent to giving the same statistical weight to all the states used as input. This way we ensure the resulting set of parameters will simultaneously handle  the spin-spin splitting in the $S$-wave, the spin-orbit and the tensor splitting, which are especially important in the $P$-wave, and the radial excitations as well.

\subsection{Diquarks}
\indent

In the study of tetraquarks, we shall treat the full four-body problem as
three  two-body problems. Repeating the steps described in the previous
subsection, we first compute the mass spectrum of the diquark, then we do the
same for the antidiquark and finally we solve the Schr\"odinger
equation once again for a two-body system composed of the diquark and antidiquark.
The inspiration for this factorization is the color structure behind it.

A diquark is a cluster of two quarks which can form a bound state. This
binding is caused by one-gluon exchange between the quarks. In this
interaction the factor $\kappa_s$ can be negative, then the short distance part
($\propto 1/r$) of the potential will be attractive. The $SU(3)$ color
symmetry of QCD implies that, when we combine two quarks in the fundamental
$(3)$ representation, we obtain:
$|qq\rangle : 3 \otimes 3 = \overline{3} \oplus 6 $.
Similarly, when we combine two antiquarks in the $\overline{3}$
representation, they can form an antidiquark in the $3$ representation.
Then the diquark and antidiquark can be combined according to
$|[qq] - [\bar{q}\bar{q}] \rangle : \overline{3} \otimes 3
= 1 \oplus 8 $ and form a color singlet, for which the one-gluon exchange
potential is also attractive (see Refs. \cite{muta,modern,Stancu} ).
The antitriplet state is attractive and yields a color factor
$\kappa_s=-2/3$, while the sextet is repulsive and yields a color
factor $\kappa_s=+1/3$ \cite{Grif:Particles,master}. Therefore
we will consider only diquarks in the antitriplet color state.
Indeed, for the single-flavor tetraquarks only the antitriplet
diquarks can build pure states \cite{WuT4c}, while the sextet diquarks
 would necessarily appear
mixed and in just a few cases. In Refs. \cite{Brink,SuHongLee} the sextet
contribution was found to
be negligible in heavy tetraquarks with different flavor structure, like $ud\bar{b}\bar{b}$.
Nevertheless, at the end of
the presentation of our results, we will present and discuss results obtained with $6 - \bar{6}$
configurations.
We will use a diquark $[c c]$ in the ground state, with no orbital
nor radial excitations, such that we have the most compact diquark. We choose the attractive antitriplet color state, which is antisymmetric in the color wavefunction. Then, in order to respect the Pauli principle (the two quarks of the same flavor are identical fermions), the diquark total spin $S$  must be 1. In this way the total wavefunction of the diquark will be antisymmetric.

Notice that going from the color factor $-4/3$ (for quark-antiquark
in the singlet color state) to the color factor $-2/3$ (for quark-quark
in the antitriplet color state) is equivalent to introducing a factor $1/2$,
which one would expect to be a global factor since it comes from the color
structure of the wavefunction. Because of that, it is very common to extend
this factor $1/2$ to the whole potential describing the quark-quark
interaction.
This rule is motivated by the interactions inside baryons, where two
quarks can also be considered to form a color-antitriplet diquark,
which can then interact with the third quark and form a color-singlet
baryon. Since this seems to give satisfactory results in baryon
spectroscopy, it has also been extended to diquarks inside tetraquarks.
The general rule would be simply $V_{qq} = V_{q\bar{q}}/2$. Many authors
with different tetraquark models, for instance
Refs. \cite{LuchaT4Q,ChinesesT4Q}, also divide the confining part of the
potential by 2 in order to adapt it to the diquark case. In our model,
besides the change in the color factor, the string tension $b$, obtained
from the fit of $c \bar{c}$ spectra, will be also divided by 2.

The calculation of the total mass of the diquark is completely analogous
to the $c\bar{c}$ mesons, as in Eq. (\ref{eq.charmonium}). The
spin-dependent corrections are also analogous since we are still dealing
with a two-body system composed of two spin 1/2 particles.

\subsection{Tetraquarks}
\indent
The all-charm tetraquark will be treated as a two-body
($c c$ - $\bar{c} \bar{c}$) system with $m_{cc} = m_{\bar{c}\bar{c}}$.
The color factor should correspond to
the color singlet, therefore we will use $\kappa_s=-4/3$ and also the
same parameters $\alpha_s$, $b$ and $\sigma$ obtained from the fit of
the $c\bar{c}$ spectrum.
The calculation of its total mass will also be analogous to the charmonium case:
\begin{equation}
    M(T_{4c}) = m_{cc} + m_{\bar{c}\bar{c}} + E_{[cc][\bar{c}\bar{c}]} \;
+\; \langle V^{(1)}_{Spin} \rangle_{[cc][\bar{c}\bar{c}]}
\end{equation}
In order to properly calculate the spin-dependent corrections we need to
remember that the diquarks have spin 1. Then, for the coupling
of a spin 1 diquark and spin 1 antidiquark, we will have the total tetraquark
spin  $S_T = 0,\;1,\;2$. Besides that, we will also allow  radial and/or
orbital excitations in the diquark-antidiquark system. In our
non-relativistic approach, we use ordinary quantum mechanics to couple the
total spin $S_T$ to the orbital angular momentum $L_T$ into the total
angular momentum $J_T$.

For the spin-spin and spin-orbit corrections, we can obtain the angular
factors from the spin, orbital and total angular momentum quantum numbers.
However, for the tensor correction we only have a general result (in terms
of eigenvalues) for the interaction between two spin 1/2 particles, shown
in Eq. \eqref{tensorS1/2}. Then, for a proper treatment of the tensor
interaction in the diquark-antidiquark system we will explicitly apply
the tensor operator on the angular part of the tetraquark wavefunction,
as we will describe below.

\begin{figure}[htb!]
\begin{center}
\includegraphics[width=0.5\textwidth]{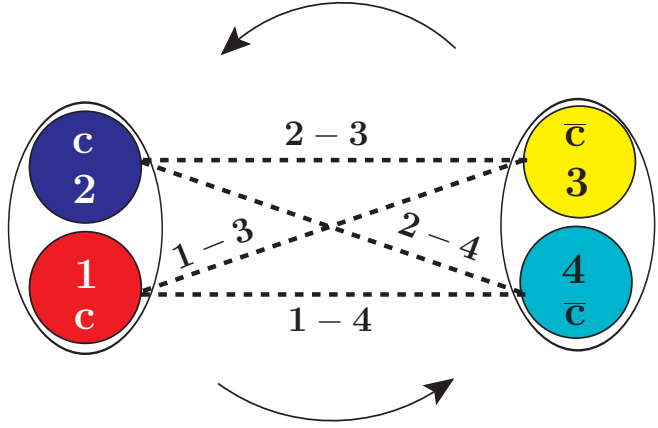} 
\caption{Pictorial representation of the tensor interaction between diquark
and antidiquark. The arrows represent the orbital angular momentum.}
\label{fig:T4cTensor}
\end{center}
\end{figure}

Let us focus on the spatial and spin components of the wavefunction.
We factorize the radial wavefunction from the angular one that combines
orbital angular momentum  and spin, which are coupled using Clebsh-Gordan
coefficients. We will use the indices 1 and 2 for the two quarks inside
the diquark, and 3 and 4 for the two antiquarks inside the antidiquark (see
Fig. \ref{fig:T4cTensor}).
To illustrate our procedure to treat tensor interactions,
we present one specific example with total spin $S_T=2$, $L_T=1$ and
$J_T=2$. $S_d$ and $S_{\bar{d}}$ will denote the total spin of the
diquark and antidiquark, respectively. We write the possible couplings
in a generic form $|S, M_S\rangle$, where $S$ is the total spin and $M_S$
is its z-component. The arrows denote the spins of each constituent, in
the order 1, 2 for the diquark and 3, 4 for the antidiquark. As usual the
up arrow denotes spin up, $|\frac{1}{2}, \frac{1}{2}\rangle$, and the down arrow denotes spin down, $|\frac{1}{2}, -\frac{1}{2}\rangle$. We show it in
terms of diquark and antidiquark spin basis, and also in terms of the two
quarks and two antiquarks spin  basis (each group of four arrows is always
in the order ``1234''). These wavefunctions were inspired by the ones
presented in Refs. \cite{Brink-Gauss,Stancu,Brink,SuHongLee}, and we
generalized them to include orbital angular momentum between diquark and
antidiquark. For the choices mentioned above the wavefunction reads:
\begin{align}
\label{11212}
&\nonumber  [(S_d=1)\otimes(S_{\bar{d}}=1)\rightarrow (S_T=2)]
\otimes(L_T=1)\longrightarrow |J_T, M_{J_T}\rangle = |2,2\rangle_{J_T}\\
& \nonumber = \sqrt{\frac{2}{3}}|2,2\rangle_{S_T}\otimes|1,0\rangle_{L_T}
  -\frac{1}{\sqrt{3}}|2,1\rangle_{S_T}\otimes|1,1\rangle_{L_T} \\
&\nonumber = \sqrt{\frac{2}{3}}
\Big{(}
|1,1\rangle_{12}\otimes|1,1\rangle_{34}
\Big{)} Y_1^{0}(\theta,\varphi)\\
&\quad -\frac{1}{\sqrt{3}}
\Big{(}
\frac{1}{\sqrt{2}}|1,1\rangle_{12}\otimes|1,0\rangle_{34}
   +\frac{1}{\sqrt{2}}|1,0\rangle_{12}\otimes|1,1\rangle_{34}
\Big{)} Y_1^{1}(\theta,\varphi) \\
&\nonumber = \sqrt{\frac{2}{3}}
\Big{(}
|\uparrow\uparrow\rangle_{12}\otimes|\uparrow\uparrow\rangle_{34}
\Big{)} Y_1^{0}(\theta,\varphi)\\
&\nonumber\quad  -\frac{1}{\sqrt{3}}
\Big{(}
\frac{1}{\sqrt{2}}|\uparrow\uparrow\rangle_{12}
\otimes|\frac{\uparrow\downarrow+\downarrow\uparrow}{\sqrt{2}}\rangle_{34}
   +\frac{1}{\sqrt{2}}|\frac{\uparrow\downarrow+\downarrow\uparrow}
{\sqrt{2}}\rangle_{12}
   \otimes|\uparrow\uparrow\rangle_{34}
\Big{)} Y_1^{1}(\theta,\varphi) \\
&\nonumber= \sqrt{\frac{2}{3}}
\Big{(}
\uparrow\uparrow\uparrow\uparrow
\Big{)} Y_1^{0}(\theta,\varphi)
-\frac{1}{\sqrt{3}}
\Big{(}
\frac{1}{2}(\uparrow\uparrow\uparrow\downarrow
+\uparrow\uparrow\downarrow\uparrow
+\uparrow\downarrow\uparrow\uparrow
+\downarrow\uparrow\uparrow\uparrow
)
\Big{)} Y_1^{1}(\theta,\varphi)
\end{align}
\vspace{5pt}
For $L_T=1$ we have seven combinations to get $J_T$ if we are considering
spin 1 diquark and antiquark: one for $S_T=0$ ($J_T=1$), three for $S_T=1$
($J_T=0,\;1,\;2$) and three for $S_T=2$ ($J_T=1,\;2,\;3$).

We now explicitly apply the tensor operator on the above angular wavefunction
and we note that within our approximations, it is equivalent to apply this
operator directly on the diquark-antidiquark pair (in spin 1 basis) or
consider  a sum of four tensor interactions between each quark-antiquark pair (spin 1/2
basis) as illustrated in Fig. \ref{fig:T4cTensor}, as would be expected from
the angular momentum algebra \footnote{To see this, we could
write $\mathbf{S_d} =  \mathbf{S_1} + \mathbf{S_2}$, $\mathbf{S_{\bar{d}}}
= \mathbf{S_3} + \mathbf{S_4}$ and open the tensor between
diquark-antidiquark into four tensor operators between quark-antiquark
pairs.}. We have:
\begin{equation}\label{Sdd}
  \mathbf{S_{d-\bar{d}}} =
   12\left(
    \frac{(\mathbf{S_d \cdot r})(\mathbf{S_{\bar{d}}\cdot r})}{\mathbf{r}^2}
     - \frac{1}{3}(\mathbf{S_d} \cdot \mathbf{S_{\bar{d}}})
    \right)
=\mathbf{S_{14}} + \mathbf{S_{13}} + \mathbf{S_{24}} + \mathbf{S_{23}}
\end{equation}
Since the tetraquark is treated as a two-body system,
the expectation value of the radial wavefunction between every
$q\bar{q}$ pair is the same and can be factorized.
In a four-body problem (using Jacobi coordinates for example) where all
the four constituents are allowed to move and interact with each other at the same time, this would not be true. This type of approach can be
found in other models of tetraquarks, for instance in
Refs.~\cite{Brink,SuHongLee}. Usually in this kind of approach only
the ground state is considered, with no orbital excitations, and hence
only the spin-spin interaction is relevant
since the spin-orbit and tensor vanish for $\ell=0$. Besides, in order to
tackle the four-body problem one needs to resort to a variational
approximation with Gaussian trial wavefunctions or similar methods, therefore
there will always be a compromise between the precision of the numerical
solution and the reliability of the assumptions.

In order to deal with the generalization of the tensor interaction to
the tetraquark case, we will rewrite the tensor in a form that allows us to
recover the same results that we already know for the particular case of two
spin 1/2 particles and that can also be used as a generalization to other
cases, such as the interaction between two spin 1 diquarks.
The operator $\mathbf{S_{12}}$ in Eq. \eqref{TensorFactor} is a ``rank-2''
tensor which can be written in terms of spin operators and spherical
harmonics, as shown in textbooks
\cite{Cohen}. 
An extensive discussion of this approach can be found in Ref. \cite{master}.

The following functional form does not use any particular relation or
eigenvalues for spin 1/2 particles, only general properties of angular
momentum elementary theory. One can write the unity vector
$\mathbf{\hat{r}}$ in spherical coordinates and the spin operators in
Cartesian components. Then they can be rearranged into raising, lowering
and z-component spin operators and spherical harmonics of $\ell=2$, and we
can write:
\begin{equation}\label{WT}
  \mathbf{S_{12}} = 4[T_0 + T_0' + T_1 + T_{-1} + T_2 + T_{-2}]
\end{equation}
where
\vspace{-10pt}
\begin{align}\label{T-Y}
\begin{aligned}
&T_0&=&\; 2\sqrt{\frac{4\pi}{5}} \; Y_{2}^{0}(\theta,\phi) \; S_{1z}S_{2z} \\
&T_0'&=&\; -\frac{1}{4}2\sqrt{\frac{4\pi}{5}} \; Y_{2}^{0}(\theta,\phi) \;
(S_{1+}S_{2-}+S_{1-}S_{2+}) \\
&T_1&=&\; \frac{3}{2}\sqrt{\frac{8\pi}{15}} \; Y_{2}^{-1}(\theta,\phi)  \;
(S_{1z}S_{2+}+S_{1+}S_{2z}) \\
&T_{-1}&=&\; -\frac{3}{2}\sqrt{\frac{8\pi}{15}} \; Y_{2}^{1}(\theta,\phi)  \;
(S_{1z}S_{2-}+S_{1-}S_{2z}) \\
&T_2&=&\; 3\sqrt{\frac{2\pi}{15}} \; Y_{2}^{-2}(\theta,\phi)  \; S_{1+}S_{2+} \\
&T_{-2}&=&\; 3\sqrt{\frac{2\pi}{15}} \; Y_{2}^{2}(\theta,\phi)  \; S_{1-}S_{2-}
\end{aligned}
\end{align}
With the expressions above we can take the expectation value of the tensor
operator in the angular wavefunctions, as in Eq.~\eqref{11212}, and use the selection rules
of the spherical harmonics to find the non-vanishing terms.

To close this subsection,  we discuss the tetraquark quantum numbers, as in
Refs. \cite{maiani1,Maiani-NS}. We can use the diquark-antidiquark basis to label the
possible quantum numbers $J^{PC}$ of the tetraquark. We shall use the following
notation:
\begin{equation}\label{T4cBase}
  |T_{4Q}\big{>} = |S_{d},S_{\bar{d}},S_T,L_T\big{>}_{J_T}
\end{equation}
where $S_{d}$ is the total spin of the diquark, $S_{\bar{d}}$ is the total spin
of the antidiquark, $S_T$ is the total spin of the tetraquark, assumed to come
from the coupling $S_{d}\otimes S_{\bar{d}}$, $L_T$ is the orbital angular momentum
relative to the diquark-antidiquark system  (in the two-body approximation), and
$J_T$ is the total angular momentum of the tetraquark, assumed to come from the
coupling $S_T\otimes L_T$.
The general formulae for charge-conjugation and parity of the tetraquark are:
\begin{align}\label{T4Q-C}
\begin{aligned}
  C_T &= (-1)^{L_T+S_T}\\
  P_T &= (-1)^{L_T}
\end{aligned}
\end{align}
Since we are interested in the $T_{4c}$ tetraquark, where the diquarks are composed
of two
charm quarks with spin 1 in the antitriplet color configuration, for the $S$-wave
states we
have the following possibilities:
\begin{align}
\begin{aligned}
  |0^{++}\big{>}_{T4c} &= |S_{cc}=1,S_{\bar{c}\bar{c}}=1,S_T=0,L_T=0\big{>}_{J_T=0}\\
  |1^{+-}\big{>}_{T4c} &= |S_{cc}=1,S_{\bar{c}\bar{c}}=1,S_T=1,L_T=0\big{>}_{J_T=1}\\
  |2^{++}\big{>}_{T4c} &= |S_{cc}=1,S_{\bar{c}\bar{c}}=1,S_T=2,L_T=0\big{>}_{J_T=2}
\end{aligned}
\end{align}
Note that all the $S$-wave tetraquarks described above have positive parity. The
introduction of
the first orbital excitation will bring a factor $(-1)$ in both parity and charge conjugation.
Then all the $P$-wave states (with $L_T=1$) will have odd parity and the opposite
charge
conjugation in comparison with the $S$-wave states. In Table \ref{tab:T4cJPC}
we list
the $J^{PC}$ quantum numbers of the 10 possibilities which we consider for the
$S$-wave and
$P$-wave  all-charm tetraquarks built with spin 1 diquarks (also in accordance with
Refs. \cite{mais2} and \cite{Chao}).
\begin{table}[H]
\centering
\renewcommand{\arraystretch}{1.3}
\caption{Results for the $J^{PC}$ quantum numbers of the $T_{4c}$
with $[S_d=S_{\bar{d}}=1
\rightarrow S_T=0,1,2]\otimes L_T=0,1$.}\vspace{10pt}
\begin{tabular}{|c|c|c|c|}
  \hline \hline
  \label{tab:T4cJPC}
  $S_T$ & $L_T$ & $J_T$ & $J^{PC}$ \\ \hline
  0 & 0 & 0 & $0^{++}$ \\
  1 & 0 & 1 & $1^{+-}$ \\
  2 & 0 & 2 & $2^{++}$ \\ \hline
  0 & 1 & 1 & $1^{--}$ \\ \hline
  1 & 1 & 2 & $2^{-+}$ \\
  1 & 1 & 1 & $1^{-+}$ \\
  1 & 1 & 0 & $0^{-+}$ \\ \hline
  2 & 1 & 3 & $3^{--}$ \\
  2 & 1 & 2 & $2^{--}$ \\
  2 & 1 & 1 & $1^{--}$ \\
  \hline \hline
\end{tabular}
\end{table}

\section{Results}
\label{Results}
\indent

In this section we present the results of the calculations with the formalism
outlined in the previous section.
We use the following notation in our  tables: the principal quantum number
is $N$ ($N=1$ for the ground state, $N=2$ for the first radial excitation and so on),
$\ell$ is the orbital angular momentum, $S$ is the total spin and $J$ the total
angular momentum.
In spectroscopy notation the states are usually labeled by $N^{2S+1}\ell_J$, with
$\ell=0,1,2,3,\,\dots \,\rightarrow\, S,\,P,\,D,\,F,\,\dots$, for example
$1^3S_1$ for $J/\psi$.

\subsection{Charmonium}
\label{Charmonium}
  \indent
\renewcommand{\arraystretch}{1.5}

\begin{figure}[H]
  \centering
  \includegraphics[width=0.45\textwidth]{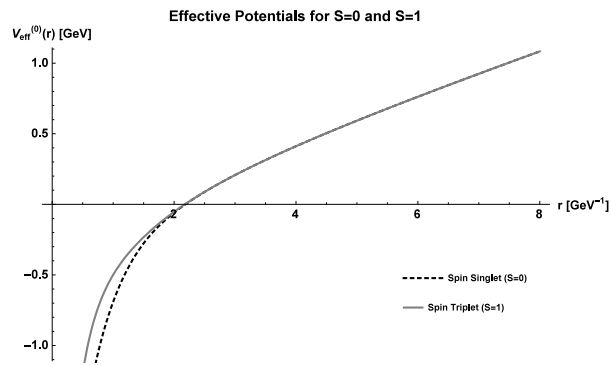}\\
  \caption{Effective Potentials:
\textit{Coulomb plus linear plus smeared spin-spin}, for $S=0$ and $S=1$.
Parameters are $\alpha_s = 0.5202$, $b=0.1463$ GeV$^2$, $\sigma=1.0831$ GeV.}
\label{PotMod2}
\end{figure}

In order to get good estimates of diquark and tetraquark properties,
we first study the spectrum of the conventional charmonium states to
observe how well we can fit the experimental data. In our model we
considered the zeroth-order potential of the
form \textit{Coulomb plus linear plus smeared spin-spin interactions}.
We
separate the spin triplet (S=1) and spin singlet (S=0) before solving the
Schr\"{o}dinger equation. Using $\kappa_s= -4/3$, $S_1=S_2=1/2$ and $S=0$
or $S=1$, we replace the operator $\mathbf{S_1 \cdot S_2}$ by the constant
$[S(S+1)-S_1(S_1+1) - S_2(S_2+1)]/2$ and we find the wavefunction $y(r)$
and the eigenvalue $E$. In Fig. \ref{PotMod2} we show the zeroth-order
potential for total spin 0 or 1. Later the spin-orbit and tensor corrections are included, splitting orbitally-excited states.
We performed a fit with experimental values from the PDG {\cite{PDG2016}. The four parameters were allowed to vary in the following range:
$1.1 < m_c < 1.9\;\mathrm{GeV},\quad 0.1 < \alpha_s < 0.7,\quad 0.050 < b < 0.450\
;\mathrm{GeV}^2,\quad 0.7<\sigma<1.3\;\mathrm{GeV}$. The results are also very
similar to those from Refs. \cite{T-Barnes,High}, which were obtained with the
fit of 11 $c\bar{c}$ states with equal statistical weight. We have included two
more, $h_c(1P)$ and $\chi_{c2}(2P)$, in a total of 13 states as input, obtaining
the following values:
\begin{equation}
m_c = 1.4622\;\mathrm{GeV},\quad \alpha_s = 0.5202, \quad b=0.1463\;
\mathrm{GeV}^2,\quad \sigma=1.0831\;\mathrm{GeV}
\end{equation}
Several fits with different number of input states and alternative models were tested in Ref. \cite{master}. 
There is one particular alternative case worth mentioning. In this case, we considered
the spin-spin interaction as first-order perturbation, proportional to
the wavefunction at the origin, with the radial coefficient given by Eq. (\ref{C:spin-spin}) (without the Gaussian smearing), and also removed the Thomas precession term
from the spin-orbit interaction, which is proportional to the string tension
$b$ on Eq. \eqref{C:spin-orbit}. In this way the spin-dependent corrections come
exclusively from the Breit-Fermi Hamiltonian describing one-gluon exchange, as
in Ref. \cite{beyond}. In this scheme it was possible to fit very
accurately the 6 ground states $1S$ and $1P$: $\eta_c (1^1S_0),\;
J/\psi (1^3S_1),\;
h_c (1^1P_1),\; \chi_{c0} (1^3P_0),\; \chi_{c1} (1^3P_1),\;
\chi_{c2} (1^3P_2)$,
with the parameter set $m_c = 1.2819$ GeV, $\alpha_s=0.3289$ and $b=0.2150$
GeV$^2$.  This set is appealing since the mass of the charm quark is exactly
the PDG value  \cite{PDG2016} obtained in the $\overline{MS}$ scheme:
$1.28 \pm 0.03$ GeV, and the
coupling constant $\alpha_s$ is also smaller, favoring the assumption of the
perturbative regime of QCD. However, for radial excitations, especially above
the $D\bar{D}$ threshold, this scheme does not work very well and hence we
restrict ourselves to present the results obtained with the model that gives
the best agreement with the whole experimental data set, since we believe this might
yield better predictions to higher new charmonium states and also for the
diquark and tetraquark.
\begin{table}[htb!]
\caption{\label{tab:charmMod2-13}
Results for charmonium $c\bar{c}$ wavefunctions
from the model. Parameters are $m_c = 1.4622$ GeV, $\alpha_s = 0.5202$,
$b=0.1463$ GeV$^2$, $\sigma=1.0831$ GeV.}
\begin{center}
\centering
\renewcommand{\arraystretch}{1.5}
\begin{tabular}{c r r r r}
\hline\hline
$N^{2S+1}\ell$ & $M^{(0)}$ [GeV] & $|R(0)|^2$ [GeV$^3$]& $\langle r^2
\rangle^{1/2}$ [fm] & $\displaystyle\left\langle \frac{v^2 }{c^2}\right\rangle$\\
\hline
    $1^1S$ & $2.9924$ & $1.5405$ & $0.375$ & $0.336$ \\
    $1^3S$ & $3.0917$ & $1.1861$ & $0.421$ & $0.253$ \\
\hline
    $1^1P$ & $3.5105$ & $0$ & $0.678$ & $0.257$ \\
    $1^3P$ & $3.5191$ & $0$ & $0.689$ & $0.246$ \\
\hline
    $2^1S$ & $3.6317$ & $0.7541$ & $0.839$ & $0.308$ \\
    $2^3S$ & $3.6714$ & $0.7092$ & $0.867$ & $0.293$ \\
\hline
    $1^1D$ & $3.7951$ & $0$ & $0.899$ & $0.280$ \\
    $1^3D$ & $3.7958$ & $0$ & $0.901$ & $0.278$ \\
\hline
    $2^1P$ & $3.9334$ & $0$ & $1.071$ & $0.324$ \\
    $2^3P$ & $3.9427$ & $0$ & $1.082$ & $0.315$ \\
\hline
    $3^1S$ & $4.0481$ & $0.6088$ & $1.210$ & $0.364$ \\
    $3^3S$ & $4.0755$ & $0.5914$ & $1.230$ & $0.357$ \\
\hline
    $2^1D$ & $4.1591$ & $0$ & $1.258$ & $0.350$ \\
    $2^3D$ & $4.1604$ & $0$ & $1.261$ & $0.348$ \\
\hline
    $4^1S$ & $4.3933$ & $0.5430$ & $1.531$ & $0.424$ \\
    $4^3S$ & $4.4150$ & $0.5340$ & $1.547$ & $0.419$ \\
\hline\hline
\end{tabular}
\end{center}
\end{table}

In Table \ref{tab:charmMod2-13} we present the wavefunction properties. Notice
that the inclusion of the spin-spin interaction in the zeroth-order potential
creates a small difference between the wavefunction of spin singlet and spin
triplet. The spin 0 states receive a negative contribution from this
interaction
in the potential, what causes the short-distance region of the potential
(small $r$ coordinate) to be ``more negative'' generating states with smaller
root mean square radius, higher value of the wavefunction at the origin and
higher quark velocity.

The spin-dependent interactions are very important in charmonium
spectroscopy because they can test  the QCD dynamics
in the heavy quark context, lying between the perturbative and the
non-perturbative regime.  Particularly interesting  is the role of the
spin-spin interaction in orbitally-excited states. It is convenient to
define the spin-average mass of a multiplet (spin here means $J$), also know
as ``center-of-weight'' or  ``center-of-gravity'' (c.o.g.):
\begin{equation}\label{cog}
 \left<M(N^{2S+1}\ell_J)\right> = \displaystyle \frac{\displaystyle\sum_{J}
(2J+1)M(N^{2S+1}\ell_J)}{\displaystyle\sum_{J} (2J+1)}, \quad\quad
\mathrm{(c.o.g.)}
\end{equation}
For $P$-wave ground state, for example, we have:
\begin{equation}\label{cogP}
   \left<M(1^3P_J)\right> = \displaystyle \frac{5M(1^3P_2)+3M(1^3P_1) +
M(1^3P_0)}{9}
\end{equation}


\begin{table}[htb!]
\caption{\label{tab:charmMod2-13Spin} Results for charmonium $c\bar{c}$ masses
from the model. Parameters are $m_c = 1.4622$ GeV, $\alpha_s = 0.5202$,
$b=0.1463$ GeV$^2$, $\sigma=1.0831$ GeV.}
\begin{center}
\centering
\renewcommand{\arraystretch}{1.5}
\begin{tabular}{c r r r r r r r r r}
\hline\hline
$N^{2S+1}\ell_J$ & $\langle T \rangle$ & $\langle V^{(0)}_{V} \rangle$ & $\langle V^{(0)}_{S} \rangle$ & $\langle V^{(0)}_{SS} \rangle$ & $E^{(0)}$ & $M^{(0)}$ [MeV] & $\langle V^{(1)}_{LS} \rangle$ & $\langle V^{(1)}_T \rangle$ & $M^{f}$ [MeV] \\
\hline
    $1^1S_0$ & $491.9$ & $-584.4$ & $246.2$ & $-85.6$ & $68.1$ & $2992.4$ & $0$ & $0$ & $2992.4$ \\
    $1^3S_1$ & $370.6$ & $-504.0$ & $279.4$ & $21.4$ & $167.4$ & $3091.7$ & $0$ & $0$ & $3091.7$ \\
\hline
    $1^3P_0$ & $359.5$ & $-246.6$ & $480.0$ & $2.0$ & $594.8$ & $3519.1$ & $-63.9$ & $-29.4$ & $3425.8$ \\
    $1^3P_1$ & $359.5$ & $-246.6$ & $480.0$ & $2.0$ & $594.8$ & $3519.1$ & $-32.0$ & $14.7$ & $3501.8$ \\
    $1^1P_1$ & $375.2$ & $-253.1$ & $471.1$ & $-7.0$ & $586.2$ & $3510.5$ & $0$ & $0$ & $3510.5$ \\
    $1^3P_2$ & $359.5$ & $-246.6$ & $480.0$ & $2.0$ & $594.8$ & $3519.1$ & $32.0$ & $-2.9$ & $3548.1$ \\
\hline
    $2^1S_0$ & $450.6$ & $-287.3$ & $573.8$ & $-29.7$ & $707.4$ & $3631.7$ & $0$ & $0$ & $3631.7$ \\
    $2^3S_1$ & $428.5$ & $-281.7$ & $590.4$ & $9.8$ & $747.1$ & $3671.4$ & $0$ & $0$ & $3671.4$ \\
\hline
    $1^3D_1$ & $407.0$ & $-175.4$ & $639.7$ & $0.2$ & $871.5$ & $3795.8$ & $-8.8$ & $-3.9$ & $3783.1$ \\
    $1^3D_2$ & $407.0$ & $-175.4$ & $639.7$ & $0.2$ & $871.5$ & $3795.8$ & $-2.9$ & $3.9$ & $3796.7$ \\
    $1^1D_2$ & $408.8$ & $-175.9$ & $638.5$ & $-0.6$ & $870.8$ & $3795.1$ & $0$ & $0$ & $3795.1$ \\
    $1^3D_3$ & $407.0$ & $-175.4$ & $639.7$ & $0.2$ & $871.5$ & $3795.8$ & $5.9$ & $-1.1$ & $3800.6$ \\
\hline
    $2^3P_0$ & $460.4$ & $-186.2$ & $742.1$ & $2.2$ & $1018.4$ & $3942.7$ & $-59.9$ & $-26.1$ & $3856.7$ \\
    $2^3P_1$ & $460.4$ & $-186.2$ & $742.1$ & $2.2$ & $1018.4$ & $3942.7$ & $-29.9$ & $13.0$ & $3925.8$ \\
    $2^1P_1$ & $474.4$ & $-190.8$ & $733.1$ & $-7.5$ & $1009.1$ & $3933.4$ & $0$ & $0$ & $3933.4$ \\
    $2^3P_2$ & $460.4$ & $-186.2$ & $742.1$ & $2.2$ & $1018.4$ & $3942.7$ & $29.9$ & $-2.6$ & $3970.0$ \\
\hline
    $3^1S_0$ & $532.8$ & $-215.4$ & $826.5$ & $-20.1$ & $1123.8$ & $4048.1$ & $0$ & $0$ & $4048.1$ \\
    $3^3S_1$ & $521.9$ & $-215.3$ & $837.7$ & $6.9$ & $1151.2$ & $4075.5$ & $0$ & $0$ & $4075.5$ \\
\hline
    $2^3D_1$ & $508.6$ & $-145.8$ & $873.0$ & $0.3$ & $1236.1$ & $4160.4$ & $-11.6$ & $-3.7$ & $4145.1$ \\
    $2^3D_2$ & $508.6$ & $-145.8$ & $873.0$ & $0.3$ & $1236.1$ & $4160.4$ & $-3.9$ & $3.7$ & $4160.2$ \\
    $2^1D_2$ & $511.3$ & $-146.5$ & $871.0$ & $-1.0$ & $1234.8$ & $4159.1$ & $0$ & $0$ & $4159.1$ \\
    $2^3D_3$ & $508.6$ & $-145.8$ & $873.0$ & $0.3$ & $1236.1$ & $4160.4$ & $7.7$ & $-1.1$ & $4167.1$ \\
\hline
    $4^1S_0$ & $620.4$ & $-179.5$ & $1044.0$ & $-15.8$ & $1469.0$ & $4393.3$ & $0$ & $0$ & $4393.3$ \\
    $4^3S_1$ & $613.2$ & $-180.6$ & $1053.0$ & $5.6$ & $1490.7$ & $4415.0$ & $0$ & $0$ & $4415.0$ \\
\hline\hline
\end{tabular}
\end{center}
\end{table}
Interestingly, in the spin average mass   the spin-orbit and
tensor corrections cancel each other and hence if the spin-spin correction
is zero in the orbitally-excited singlet state ($1^1P_1$ for instance), its
mass should be equal to this
spin average. However, the spin-spin correction is zero
for orbitally-excited states only if it is treated as a first-order perturbation
proportional to the wavefunction at the origin. In our model, where we
include the Gaussian term non-perturbatively, there will be a small difference.
Therefore, in the present model the value of the mass $M^{(0)}$ (before the splitting due to the orbital and tensor spin-dependent corrections) of the orbitally-excited states with total spin $S=1$, like the $1^3P$, is equal to the c.o.g. of these states.

In Table \ref{tab:charmMod2-13Spin} we present the results for the masses
including the spin interactions. Note that the contribution of the spin-spin interaction to orbitally-excited states is not zero, especially in $P$-wave, even tough the wave function at the origin is still compatible with zero. Because of that the spin singlet in orbitally-excited states is slightly different from the spin-average (c.o.g.). The experimental measurements of  $1P$ states suggest that they should be very close (see Table \ref{tab:charmExp} for experimental values).
As pointed in Ref. \cite{Zhang}, a precise measurement of the difference
between the c.o.g. of the $1^3P_J$ states and the singlet $1^1P_1$ can provide useful
information about the spin-dependent interactions in heavy quarks.
Actually, the prediction for $h_c(1^1P_1)$ is already close to the
experimental value and even more so if one considers its mass as the
spin-average of the $^3P_J$ states (as done in Ref. \cite{High} for the
calculations where its mass was required). Also, the inclusion of the
recently measured $\chi_{c2}(2P)$
\cite{Uehara:2005qd,Aubert:2010ab} did not affect much the resulting
set, even though the prediction for its mass is a little higher than the
experimental value.

\begin{table}[htb!]
\caption{\label{tab:charmExp} Comparison of charmonium $c\bar{c}$
experimental data and our results (Tab. \ref{tab:charmMod2-13Spin}). Units are MeV.}
\begin{center}
\renewcommand{\arraystretch}{1.5}
\begin{tabular}{c c c c c c l}
\hline\hline
    $N^{2S+1}\ell_J$ & $M^{f}$ & Exp. \cite{PDG2016} & Meson & $J^{PC}$ \\
\hline
    $1^1S_0$ & $2992.4$ & $2983.4\pm0.5$ & $\eta_c(1S)$ & $0^{-+}$  \\
    $1^3S_1$ & $3091.7$ & $3096.900\pm0.006$ & $J/\psi(1S)$ & $1^{--}$  \\
\hline
    $1^3P_0$ & $3425.8$ & $3414.75\pm0.31$ & $\chi_{c0}(1P)$ & $0^{++}$ \\
    $1^3P_1$ & $3501.8$ & $3510.66\pm0.07$ & $\chi_{c1}(1P)$ & $1^{++}$ \\
    $1^1P_1$ & $3510.5$ & $3525.38\pm0.11$ & $h_c(1P)\;^{\dag}$ & $1^{+-}$ \\
    $1^3P_2$ & $3548.1$ & $3556.20\pm0.09$ & $\chi_{c2}(1P)$ & $2^{++}$ \\
    $1^3P$ (c.o.g.) & $(3519.1)$ & $(3525.303)$ & $-$ & $-$ \\
\hline
    $2^1S_0$ & $3631.7$ & $3639.2\pm1.2$ & $\eta_c(2S)$ & $0^{-+}$  \\
    $2^3S_1$ & $3671.4$ & $3686.097\pm0.025$ & $\psi(2S)$ & $1^{--}$  \\
\hline
    $1^3D_1$ & $3783.1$ & $3773.13\pm0.35$ & $\psi(3770)$ &$1^{--}$ \\
    $1^3D_2$ & $3796.7$ & $-$ & $-$ & $2^{--}$ \\
    $1^1D_2$ & $3795.1$ & $-$ & $-$ & $2^{-+}$ \\
    $1^3D_3$ & $3800.6$ & $-$ & $-$ & $3^{--}$ \\
    $1^3D$ (c.o.g.) & $(3795.8)$ & $-$ & $-$ & $-$ & $-$ \\
\hline
    $2^3P_0$ & $3856.7$ &$-$ & $*$ & $0^{++}$ \\
    $2^3P_1$ & $3925.8$ & $-$ & $-$ & $1^{++}$ \\
    $2^1P_1$ & $3933.4$ & $-$ & $-$ & $1^{+-}$ \\
    $2^3P_2$ & $3970.0$ & $3927.2\pm2.6$ & $\chi_{c2}(2P)$ & $2^{++}$ \\
    $2^3P$ (c.o.g.) & $(3942.7)$ & $-$ & $-$ & $-$ \\
\hline
    $3^1S_0$ & $4048.1$ & $-$ & $-$ & $0^{-+}$  \\
    $3^3S_1$ & $4075.5$ & $4039\pm1$ & $\psi(4040)$ & $1^{--}$  \\
\hline
    $2^3D_1$ & $4145.1$ & $4191\pm5$ & $\psi(4160)$ & $1^{--}$ \\
    $2^3D_2$ & $4160.2$ & $-$ & $-$ & $2^{--}$ \\
    $2^1D_2$ & $4159.1$ & $-$ & $-$ & $2^{-+}$ \\
    $2^3D_3$ & $4167.1$ & $-$ & $-$ & $3^{--}$ \\
    $2^3D$ (c.o.g.) & $(4158.9)$ & $-$ &$-$ & $-$ \\
\hline
    $4^1S_0$ & $4393.3$ & $-$& $-$ & $0^{-+}$  \\
    $4^3S_1$ & $4415.0$ & $4421\pm4$ & $\psi(4415)$ & $1^{--}$  \\
\hline\hline
\end{tabular}
\end{center}
\rm $^{\dag}$ {\footnotesize In Ref. \cite{High} the $h_c(1P)$ is taken
as the spin-average (c.o.g.) of the $P$-wave states, which is in better
agreement with experimental data.}\\
\rm $*$ See text for discussion about the $\chi_{c0}(2P)$ and the $X(3915)$. \\
\end{table}

In the 2014 edition of the PDG \cite{PDG2014} the $X(3915)$ was assigned as
the $2^3 P_0$  $c\bar{c}$ state, the $\chi_{c0}(2P)$,
but due to many reasons \cite{OlsenX3915} it has been removed from this
position. The $X(3915)$ still has the status of an exotic resonance. A
discussion about its nature (and also about the $\chi_{c2}(2P)$ state)
can be found in Ref. \cite{Ortega:2017qmg}. A recent example of the $X(3915)$
interpreted as a diquark-antidiquark tetraquark $[cs][\bar{c}\bar{s}]$ can
be found in Ref. \cite{Lebed:2016}.
In Ref. \cite{WhereGuo} an analysis of Belle \cite{Uehara:2005qd} and BaBar
\cite{Aubert:2010ab} data showed some evidence of the ``real'' $\chi_{c0}(2P)$
indicating that its mass could be around $3837.6\pm11.5$ MeV, which is in
better agreement with quarkonium models. Recently, the Belle Collaboration
found a candidate for the $\chi_{c0}(2P)$ in the data of
$e^+e^- \to J/\psi D \bar{D}$ \cite{Chilikin:2017evr}, with a mass of
$3862^{+26+40}_{-32-13}$ MeV and a width of $201^{+154+88}_{-67-82}$ MeV.

Finally, in Table \ref{tab:charmExp} we compare the results of the model
with the experimental data, which are illustrated in the mass spectrum presented
in Fig. \ref{charmMod2-13}. We can see that the agreement with the experimental
data is satisfactory.

\begin{figure}[H]
  \centering
  \includegraphics[width=\textwidth]{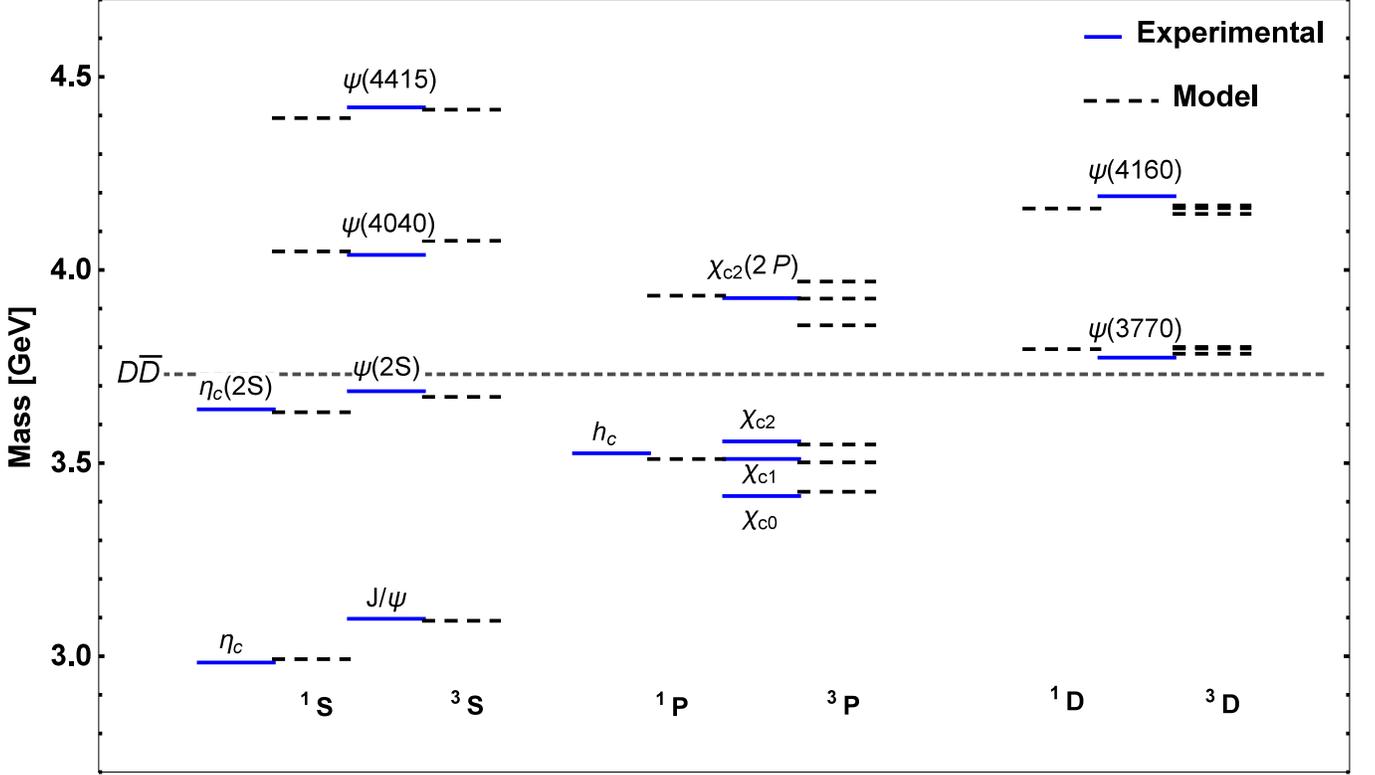}\\
 \caption{Spectrum of charmonium. Solid lines: experimental
data \cite{PDG2016}. Dashed lines: results from the model.
Parameters are $m_c = 1.4622$ GeV, $\alpha_s = 0.5202$, $b=0.1463$ GeV$^2$,
and $\sigma=1.0831$ GeV. Each state is shown with experimental data on the left
and model results on the right. Notice that for some of the calculated states
there is no experimental data to compare with.}
\label{charmMod2-13}
\end{figure}


\subsection{Diquarks}
\label{Diquarks}
  \indent
\renewcommand{\arraystretch}{1.5}

We now present our calculations for heavy diquarks composed of two charm
quarks $cc$ (which are equivalent for antidiquarks $\bar{c}\bar{c}$ in our
framework). We use the  model for charmonium discussed in the previous
subsection, except that due to the different color structure, the color factor
is now $\kappa_s=-2/3$, which corresponds to the attractive antitriplet color
state, and the string tension $b$ will be half of that obtained for the
$c\bar{c}$ charmonium mesons. We will adopt the parameter set  obtained by
fitting this model to 13 $c\bar{c}$ states.

In Tables \ref{tab:diqMod2-13} and \ref{tab:diqMod2-13Spin} we present the
results for the diquark wavefunctions and masses, respectively. For
completeness we also show diquarks in the $1P$, $2S$ and $2P$ states. Because of
the restrictions due to the Pauli exclusion principle the possibilities are
much less numerous. Also, since the $P$-wave introduces a $(-1)$ factor in the
parity,
the antisymmetric restriction in the wavefunction implies that their total
spin $S$ should be 0 if they are in the antitriplet color state.

\begin{table}[H]
\caption{\label{tab:diqMod2-13} Results for diquark $cc$ wavefunctions.
Parameters from charmonium fit: $m_c = 1.4622$ GeV, $\alpha_s = 0.5202$,
$b=b_{c\bar{c}}/2=0.1463/2$ GeV$^2$, $\sigma=1.0831$ GeV.}
\begin{center}
\centering
\begin{tabular}{c r r r r}
\hline\hline
$N^{2S+1}\ell$ & $M^{(0)}$ [GeV] & $|R(0)|^2$ [GeV$^3$]& $\langle r^2
\rangle^{1/2}$ [fm] & $\displaystyle\left\langle \frac{v^2 }{c^2}\right
\rangle$ \\
\hline
$\mathbf{1^3S}$ & $\mathbf{3.1334}$ & $\mathbf{0.3296}$ & $\mathbf{0.593}$ &
$\mathbf{0.123}$ \\
\hline
    $1^1P$ & $3.3530$ & $0$ & $0.906$ & $0.141$ \\
\hline
    $2^3S$ & $3.4560$ & $0.2370$ & $1.147$ & $0.167$ \\
\hline
    $2^1P$ & $3.6062$ & $0$ & $1.395$ & $0.190$ \\
\hline\hline
\end{tabular}
\end{center}
\end{table}

\begin{table}[H]
\caption{\label{tab:diqMod2-13Spin} Results for the $cc$ diquark.
Parameters from charmonium fit: $m_c = 1.4622$ GeV, $\alpha_s = 0.5202$,
$b=b_{c\bar{c}}/2=0.1463/2$ GeV$^2$, $\sigma=1.0831$ GeV.}
\begin{center}
\centering
\begin{tabular}{c r r r r r r r r r}
\hline\hline
$N^{2S+1}\ell_J$ & $\langle T \rangle$ & $\langle V^{(0)}_{V} \rangle$ &
$\langle V^{(0)}_{S} \rangle$ & $\langle V^{(0)}_{SS} \rangle$ & $E^{(0)}$ &
$M^{(0)}$ [MeV] & $\langle V^{(1)}_{LS} \rangle$ & $\langle V^{(1)}_T
\rangle$ & $M^{f}$ [MeV] \\
\hline
$\mathbf{1^3S_1}$ & $\mathbf{180.4}$ & $\mathbf{-173.9}$ & $\mathbf{197.9}$ &
$\mathbf{4.7}$ & $\mathbf{209.0}$ & $\mathbf{3133.4}$ & $\mathbf{0}$ &
$\mathbf{0}$ & $\mathbf{3133.4}$ \\
\hline
$1^1P_1$ & $206.7$ & $-93.3$ & $316.2$ & $-0.9$ & $428.7$ & $3353.0$ & $0$ & $0$ & $3353.0$ \\
\hline
   $2^3S_1$ & $244.8$ & $-105.7$ & $389.8$ & $2.9$ & $531.7$ & $3456.0$ & $0$ & $0$ & $3456.0$ \\
\hline
   $2^1P_1$ & $277.5$ & $-72.3$ & $477.9$ & $-1.2$ & $681.9$ & $3606.2$ & $0$ & $0$ & $3606.2$ \\
\hline\hline
\end{tabular}
\end{center}
\end{table}

In  Table \ref{tab:diqComp} we show a few results from other works about
$cc$ diquarks. Due
to differences in the models and presentation in each reference, we show only
the information that can be compared to our results. In particular, we select
only the results that correspond to the (attractive) antitriplet-color
configuration.  As can be seen, the $1S$ diquark is very similar in all
the models, with a mass around $3.1$ GeV.

\begin{table}[H]
\caption{\label{tab:diqComp} Results for  $cc$ diquarks from other works.}
\begin{center}
\centering
\begin{tabular}{c c c c c}
\hline\hline
$N\ell$ & $M_{cc}$ [GeV] & $|R(0)|^2$ [GeV$^3$]& $\langle r^2 \rangle^{1/2}$
[fm] & Ref. \\
\hline
$\mathbf{1S}$ & $\mathbf{3.13}$ & $\mathbf{(0.523)^2=0.2735}$ & $\mathbf{0.58}$
&     \cite{russos-BARION} \\
    $2S$ & $3.47$ & $(0.424)^2=0.1798$ & $1.12$ & \cite{russos-BARION} \\
    $2P$ & $3.35$ & $-$ & $0.88$ & \cite{russos-BARION} \\
\hline
    $\mathbf{1S}$ & $\mathbf{3.226}$ & $-$ & $-$ & \cite{LuchaT4Q} \\
\hline
  $\mathbf{1S}$ & $\mathbf{3.067}$ & $-$ & $-$ & \cite{narrow} mod. I \\
  $\mathbf{1S}$ & $\mathbf{3.082}$ & $-$ & $-$ & \cite{narrow} mod. II \\
  $1P$ & $3.523$ & $-$ & $-$ & \cite{narrow} mod. I \\
  $1P$ & $3.513$ & $-$ & $-$ & \cite{narrow} mod. II \\
\hline
$\mathbf{1S}$ & $\mathbf{3.204}$ & $-$ & $-$ & \cite{Karliner}\\
\hline\hline
\end{tabular}
\end{center}
\end{table}

\subsection{Tetraquarks}
\label{Tetraquark}
  \indent
\renewcommand{\arraystretch}{1.5}

As discussed above,  the diquark-antidiquark tetraquark is treated as a
two-body system. The diquark masses were presented in the previous
subsection and  the  parameter set was obtained from a fit to
the charmonium data. The tetraquark spectrum is calculated by replacing the charm quark mass by the diquark
mass $m_{cc}$.

We now present  the spectrum of the all-charm tetraquark considering the ground
states $1S$ and the first orbital excitations $1P$ (relative to the
diquark-antidiquark system), including all the possible combinations of total
spin and total angular momentum. We also include the radial excitations $2S$
and $2P$, in a total of 20 $T_{4c}$ states
built with two $cc$ diquarks, each of them being in an antitriplet color
state and spin 1 ($1^3S_1$).
These 20  states were built considering the coupling of the
total spin of the tetraquark $S_T$ (composed of the coupling of the total
spins of the diquark $S_d$ and antidiquark $S_{\bar{d}}$) with the relative
orbital angular momentum $L_T$ between diquark and antidiquark, resulting
in a total angular momentum $J_T$ of the tetraquark, in analogy to the
$c\bar{c}$ charmonium spectrum. The corresponding parity and charge-conjugation
quantum numbers of each combination are
compiled in Table \ref{tab:T4cJPC}.

\begin{figure}[htb!]
  \centering
  \includegraphics[width=\textwidth]{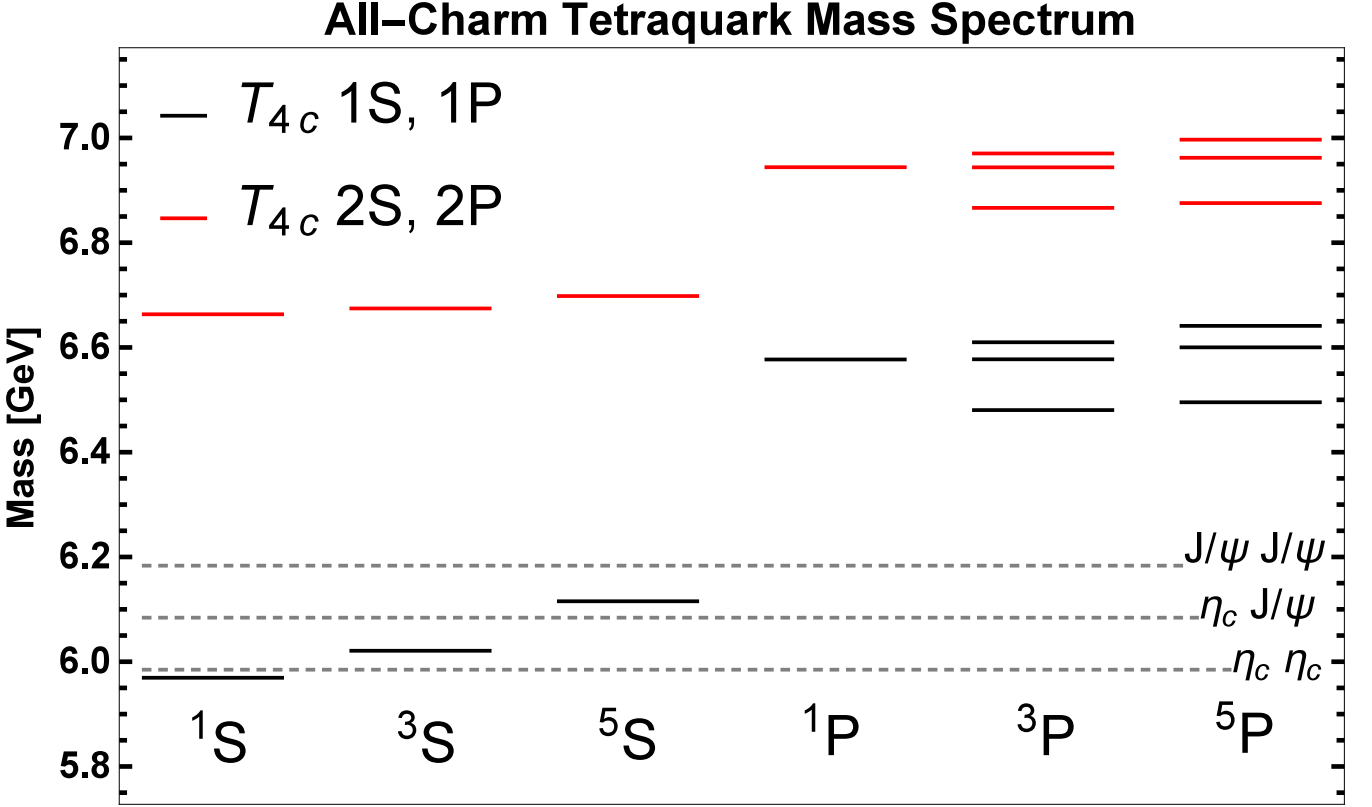}\\
 \caption{Spectrum of $T_{4c}$ obtained with the model, using ground
state ($1^3S_1$) diquark and antidiquark. Parameters are $m_{cc} = 3133.4$ MeV,
$\alpha_s = 0.5202$, $b=0.1463$ GeV$^2$, $\sigma=1.0831$ GeV.}
\label{fig:T4cMod2-13}
\end{figure}

In our model the spin-spin interaction is treated non-perturbatively. In
mesons and diquarks we had only two possibilities for total spin when
combining two spin 1/2 particles $S=0,\,1$. Now, since we consider spin 1
diquark and antidiquark, we have three possibilities for total spin
$S_T=0,\,1,\,2$, and therefore three different zeroth-order potentials, and consequently three wavefunctions for each $NL_T$ state, as presented
in Table \ref{tab:T4cMod2-13}. The splitting structure from the perturbative
corrections (spin-orbit and tensor) also has more possibilities, as presented
in Table \ref{tab:T4cMod2-13Spin} with the masses of the 20 $T_{4c}$ states.
In Figure \ref{fig:T4cMod2-13} we show the mass spectrum.

From Table \ref{tab:T4cMod2-13} we can observe that the tetraquark is
very compact. In fact, its $\langle r^2 \rangle^{1/2}$ is even smaller
than the ground state diquark.

This result apparently invalidates our
initial assumptions, which implied a two-body diquark-antidiquark interaction where the (anti)diquarks are considered as point-like objects.
However,  the large diquark radius may just be an artifact of the Cornell-like potential
used to describe the $c c$ ($\bar{c} \bar{c}$) interaction. The  result obtained  only tells us that either the real $c c$ interaction is
not Cornell-like or that the diquark-antidiquark picture is not correct. Knowing that the diquark-antidiquark was successful in describing the
recently observed multiquark states, we rather tend to question the Cornell-like potential (which was arbitrarily chosen)
for this system. Indeed, there are calculations
indicating that perhaps the dominant short-distance interaction between two quarks (or two antiquarks) is mediated by (non-perturbative)
instantons and not (perturbative) one-gluon exchange. This interaction is attractive and strong in some channels (see for example,
Ref.~\cite{shur}). In the present work we choose to keep using the
Cornell-like potential  because we need to know the diquark mass to use it as input in the final two-body (diquark-antidiquark) problem.
One could simply take the diquark mass as a free parameter and try to adjust it. However, obtaining it from the solution of the Schr\"odinger equation
is a good strategy, at least from the practical point of
view.
The use of the Cornell potential for the quark-quark interaction is an elegant way to estimate the diquark mass taking into account the diquark color structure in analogy to the $c-\bar{c}$ interaction, which successfully describes the charmonium spectrum.
In fact, in Ref.  \cite{patel},  using this
sequence of three two-body problems allowed the authors to successfully reproduce many properties of the already measured multiquark
states.  In the future the Cornell potential should be replaced by some more
realistic quark-quark interaction. For now, we would like to use the obtained results and consider them as ``privileged'' guesses for the
diquark masses.

As suggested in Ref. \cite{Chao}, the two-body approximation is better for orbitally-excited states, such as the $P$-wave considered here,
since the centrifugal barrier would suppress overlap at the origin. As we can see in Table \ref{tab:T4cMod2-13}, the compactness of the
$T_{4c}$ is also reflected in the value of the wavefunction at the origin for the $1S$ states, which is very large.

\begin{table}[H]
\caption{\label{tab:T4cMod2-13} Results for $T_{4c}$ wavefunctions and
ground state ($1^3S_1$) diquark and antidiquark. Parameters are
$m_{cc} = 3133.4$ MeV, $\alpha_s = 0.5202$, $b=0.1463$ GeV$^2$,
$\sigma=1.0831$ GeV.}
\begin{center}
\footnotesize\rm
\centering
\begin{tabular}{c r r r r}
\hline\hline
    $N^{2S_T+1}L_T$ & $M^{(0)}$ [GeV] & $|R(0)|^2$ [GeV$^3$]& $\langle r^2 \rangle^{1/2}$ [fm] & $\displaystyle\left\langle \frac{v^2 }{c^2}\right\rangle$ \\
\hline
    $1^1S$ & $5.9694$ & $8.4219$ & $0.232$ & $0.199$ \\
    $1^3S$ & $6.0209$ & $7.8384$ & $0.241$ & $0.183$ \\
    $1^5S$ & $6.1154$ & $6.6727$ & $0.264$ & $0.153$ \\
\hline
    $1^1P$ & $6.5771$ & $0$ & $0.471$ & $0.119$ \\
    $1^3P$ & $6.5847$ & $0$ & $0.478$ & $0.115$ \\
    $1^5P$ & $6.5984$ & $0$ & $0.491$ & $0.107$ \\
\hline
    $2^1S$ & $6.6633$ & $2.8414$ & $0.588$ & $0.131$ \\
    $2^3S$ & $6.6745$ & $2.8528$ & $0.595$ & $0.130$ \\
    $2^5S$ & $6.6981$ & $2.8616$ & $0.610$ & $0.129$ \\
\hline
    $2^1P$ & $6.9441$ & $0$ & $0.785$ & $0.132$ \\
    $2^3P$ & $6.9500$ & $0$ & $0.790$ & $0.130$ \\
    $2^5P$ & $6.9610$ & $0$ & $0.800$ & $0.126$ \\
\hline\hline
\end{tabular}
\end{center}
\end{table}

\begin{table}[H]
\caption{\label{tab:T4cMod2-13Spin}  Results for $T_{4c}$ masses using
ground state ($1^3S_1$) diquarks. Parameters are $m_{cc} = 3133.4$ MeV,
$\alpha_s = 0.5202$, $b=0.1463$ GeV$^2$, and $\sigma=1.0831$ GeV.}
\begin{center}
\centering
\begin{tabular}{c r r r r r r r r r l}
\hline\hline
    $N^{2S_T+1}{L_T}_{J_T}$ & $\langle T \rangle$ & $\langle V^{(0)}_{V} \rangle$ & $\langle V^{(0)}_{S} \rangle$ & $\langle V^{(0)}_{SS} \rangle$ & $E^{(0)}$ & $M^{(0)}$ [MeV] & $\langle V^{(1)}_{LS} \rangle$ & $\langle V^{(1)}_T \rangle$ & $M^{f}$ [MeV] & $J^{PC}$ \\
\hline
    $1^1S_0$ & $624.0$ & $-966.6$ & $151.1$ & $-106.0$ & $-297.3$ & $5969.4$ & $0$ & $0$ & $5969.4$ & $0^{++}$ \\
    $1^3S_1$ & $574.8$ & $-928.0$ & $157.6$ & $-50.2$ & $-245.8$ & $6020.9$ & $0$ & $0$ & $6020.9$ & $1^{+-}$ \\
    $1^5S_2$ & $479.4$ & $-847.5$ & $172.5$ & $44.3$ & $-151.3$ & $6115.4$ & $0$ & $0$ & $6115.4$ & $2^{++}$ \\
\hline
    $1^1P_1$ & $372.6$ & $-371.8$ & $325.3$ & $-15.8$ & $310.3$ & $6577.1$ & $0$ & $0$ & $6577.1$ & $1^{--}$ \\
\hline
    $1^3P_0$ & $358.9$ & $-364.3$ & $330.7$ & $-7.4$ & $318.0$ & $6584.7$ & $-59.4$ & $-44.8$ & $6480.4$ & $0^{-+}$ \\
    $1^3P_1$ & $358.9$ & $-364.3$ & $330.7$ & $-7.4$ & $318.0$ & $6584.7$ & $-29.7$ & $22.4$ & $6577.4$ & $1^{-+}$ \\
    $1^3P_2$ & $358.9$ & $-364.3$ & $330.7$ & $-7.4$ & $318.0$ & $6584.7$ & $29.7$ & $-4.5$ & $6609.9$ & $2^{-+}$ \\
\hline
    $1^5P_1$ & $335.4$ & $-350.8$ & $340.7$ & $6.4$ & $331.7$ & $6598.4$ & $-75.9$ & $-27.2$ & $6495.4$ & $1^{--}$ \\
    $1^5P_2$ & $335.4$ & $-350.8$ & $340.7$ & $6.4$ & $331.7$ & $6598.4$ & $-25.3$ & $27.1$ & $6600.2$ & $2^{--}$ \\
    $1^5P_3$ & $335.4$ & $-350.8$ & $340.7$ & $6.4$ & $331.7$ & $6598.4$ & $50.6$ & $-7.7$ & $6641.2$ & $3^{--}$ \\
\hline
    $2^1S_0$ & $410.8$ & $-397.0$ & $404.6$ & $-21.8$ & $396.6$ & $6663.3$ & $0$ & $0$ & $6663.3$ & $0^{++}$ \\
    $2^3S_1$ & $408.7$ & $-398.2$ & $408.7$ & $-11.4$ & $407.8$ & $6674.5$ & $0$ & $0$ & $6674.5$ & $1^{+-}$ \\
    $2^5S_2$ & $403.0$ & $-400.7$ & $416.8$ & $12.3$ & $431.4$ & $6698.1$ & $0$ & $0$ & $6698.1$ & $2^{++}$ \\
\hline
    $2^1P_1$ & $414.9$ & $-262.9$ & $537.5$ & $-12.0$ & $677.4$ & $6944.1$ & $0$ & $0$ & $6944.1$ & $1^{--}$ \\
\hline
    $2^3P_0$ & $407.8$ & $-260.0$ & $541.2$ & $-5.7$ & $683.3$ & $6950.0$ & $-47.9$ & $-35.6$ & $6866.5$ & $0^{-+}$ \\
    $2^3P_1$ & $407.8$ & $-260.0$ & $541.2$ & $-5.7$ & $683.3$ & $6950.0$ & $-23.9$ & $17.8$ & $6943.9$ & $1^{-+}$ \\
    $2^3P_2$ & $407.8$ & $-260.0$ & $541.2$ & $-5.7$ & $683.3$ & $6950.0$ & $23.9$ & $-3.6$ & $6970.4$ & $2^{-+}$ \\
\hline
    $2^5P_1$ & $394.5$ & $-254.2$ & $548.7$ & $5.2$ & $694.3$ & $6961.0$ & $-63.1$ & $-22.2$ & $6875.6$ & $1^{--}$ \\
    $2^5P_2$ & $394.5$ & $-254.2$ & $548.7$ & $5.2$ & $694.3$ & $6961.0$ & $-21.0$ & $22.2$ & $6962.1$ & $2^{--}$ \\
    $2^5P_3$ & $394.5$ & $-254.2$ & $548.7$ & $5.2$ & $694.3$ & $6961.0$ & $42.1$ & $-6.3$ & $6996.7$ & $3^{--}$ \\
\hline\hline
\end{tabular}
\end{center}
\end{table}

In Table \ref{tab:T4cMod2-13Spin} we see that the compactness of the $1S$ states
is mainly caused by  the Coulomb interaction. This suggests  that the one-gluon exchange is indeed the dominant mechanism responsible for the very strong
binding between diquark and antidiquark, which causes the energy
eigenvalue $E$ to be negative. This also implies that the spin-spin
interaction is strong. In this case we must have in mind that the
factors coming from $\mathbf{S_1 \cdot S_2}$ are larger for the coupling
of two spin 1 than for two spin 1/2 particles. It is interesting to see that
even though the spin-dependent terms are now suppressed by a factor
$1/m_{cc}^2$ and one would naturally expect them to be smaller when
compared to the corresponding terms in $c\bar{c}$ mesons, the color
interaction brings diquark and antidiquark so close that the suppression
due to this factor is overwhelmed by the huge superposition at the origin
of the system. The confinement term on the other hand, increases its
contribution as radial or orbital excitations are included, as
in $c\bar{c}$ mesons.

In Fig. \ref{fig:T4cMod2-13} we can see that the masses of the 20 states are
concentrated in the range between 6 and 7 GeV.
Among the $1S$ states, the lowest one, with $J^{PC}=0^{++}$ lies very close to
the $\eta_c$ pair threshold. Within our uncertainties (both from the choice of
parameters as well as the assumption of the diquark-antidiquark structure), we
cannot say whether this state is below or above such a threshold. If it is
above, it could be seen as a narrow state in the $\eta_c\eta_c$ invariant mass.
If it is below, then it would be stable against the rearrangement in $c\bar{c}$
pairs and other mechanisms would be necessary. Several possibilities are discussed
in Ref. \cite{Karliner}, with special attention to the $0^{++}$ lowest state, such as $T_{4c} \to D \bar{D}$ through $c\bar{c} \to g \to q\bar{q}$ . On the other hand, in Ref. \cite{Chao} several decay possibilities of the orbitally-excited states are also discussed and branching fractions are estimated. It is interesting to see that even our estimates for the excited states with $NL_T=2S,\,2P$ are below the threshold of decay into doubly-charmed baryon pairs due to light quark pair creation, $cc\bar{c}\bar{c} \to (ccq)+(\bar{c}\bar{c}\bar{q})$, which is above 7 MeV.

The second lowest state, with quantum numbers $J^{PC}=1^{+-}$, could rearrange
itself
into $\eta_c J/\psi$. However, this state seems to be more than 50 MeV below this
two-meson threshold, and therefore it should be stable.
The highest $1S$ state, with quantum numbers $J^{PC}=2^{++}$, is also more than 50
MeV below the corresponding $J/\psi$ pair threshold. It could still decay into
$\eta_c$ pairs in the $D$-wave, but this mechanism should be suppressed.

In order to be consistent with our $c\bar{c}$ results, in Fig. \ref{fig:T4cMod2-13} the two-meson
thresholds are shown using the values of charmonium masses obtained with our model, which were compared
to the experimental values in Table \ref{tab:charmExp}. In Table
\ref{tab:T4cThresholds} we compare all the $1S$ and $1P$ $T_{4c}$ states with the
corresponding lowest $S$-wave two-meson thresholds. We see that while the $1S$
states lie close to or below their thresholds, the orbitally-excited ones  are close to or above the corresponding thresholds. Therefore, it would be interesting to search for these states in the two-meson invariant mass distributions, since some of them could show up as narrow peaks just around the threshold, like the $1^{--}$ state (from the $1^1P_1$ configuration) in the $\eta_c(1S)\,h_c(1P)$ invariant mass at $6.50$ GeV, the $2^{--}$ (from the $1^5P_2$ configuration) in the $J/\psi(1S)\,\chi_{c1}(1P)$ invariant mass at $6.60$ GeV, and the $3^{--}$ state (from the $1^5P_3$ configuration) in the $J/\psi(1S)\,\chi_{c2}(1P)$ invariant mass at $6.65$ GeV.

One of these orbitally-excited states is of particular interest, since it presents
exotic quantum numbers that cannot be obtained as a simple $c\bar{c}$ system: the
$1^{-+}$ (from the $1^3P_1$ configuration). This state could be searched for in
the $\eta_c(1S)\,\chi_{c1}(1P)$ invariant mass. However, it might be quite broad
since our predictions show that it is about 80 MeV above its two-meson threshold.

\begin{table}[H]
\caption{\label{tab:T4cThresholds}  Comparison of $1S$ and $1P$ $T_{4c}$
masses with lowest $S$-wave two $c\bar{c}$ meson thresholds, either calculated
with the model or from experimental values \cite{PDG2016}. Units are MeV.}
\begin{center}
\centering
\begin{tabular}{ l  c  c  c  c  c }
\hline\hline
 & & & & \multicolumn{2}{c}{$(M_1 + M_2)$ } \\
    $J^{PC}$ & $N^{2S_T+1}{L_T}_{J_T}$ & $M_{T4c}$ & $M_1 \, M_2$ & Model & Exp.   \\
\hline
$0^{++}$ & $1^1S_0$ & $5969.4$ & $\eta_c(1S)\,\eta_c(1S)$ & $5984.8$ & $5966.8$ \\
\hline
$1^{+-}$ & $1^3S_1$ & $6020.9$ & $J/\psi(1S)\,\eta_c(1S)$ & $6084.1$ & $6080.3$ \\
\hline
$2^{++}$ & $1^5S_2$ & $6115.4$ & $J/\psi(1S)\,J/\psi(1S)$ & $6183.4$ & $6193.8$ \\
\hline
$0^{-+}$ & $1^3P_0$ & $6480.4$ & $\eta_c(1S)\,\chi_{c0}(1P)$ & $6418.2$ & $6398.1$ \\
\hline
$1^{-+}$ & $1^3P_1$ & $6577.4$ & $\eta_c(1S)\,\chi_{c1}(1P)$ & $6494.2$ & $6494.1$ \\
\hline
$1^{--}$ & $1^5P_1$ & $6495.4$ & $\eta_c(1S)\,h_c(1P)$ & $6502.9$ & $6508.8$ \\
$1^{--}$ & $1^1P_1$ & $6577.1$ & $$ & $$ & $$ \\
\hline
$2^{-+}$ & $1^3P_2$ & $6609.9$ & $\eta_c(1S)\,\chi_{c2}(1P)$ & $6540.5$ & $6539.6$ \\
\hline
$2^{--}$ & $1^5P_2$ & $6600.2$ & $J/\psi(1S)\,\chi_{c1}(1P)$ & $6593.5$ & $6607.6$ \\
\hline
$3^{--}$ & $1^5P_3$ & $6641.2$ & $J/\psi(1S)\,\chi_{c2}(1P)$ & $6639.8$ & $6653.1$ \\
\hline\hline
\end{tabular}
\end{center}
\end{table}

Next, we comment on the results of other works which also investigate the
existence and properties of this state composed of four charm quarks.
Some of them also consider a sextet structure for the diquarks (which can
also lead to a color singlet tetraquark). In the following tables we present
a compilation of  the main results.

\begin{table}[H]
\caption{\label{tab:T4cCompNarrow} Results for the $T_{4c}$ mass (without
spin-corrections) from Ref. \cite{narrow}.}
\begin{center}
\footnotesize\rm
\centering
\begin{tabular}{c c c c}
\hline\hline
$NL_T$ & $M^{(0)}_{T4c}$ [GeV] & Model & Color\\
\hline
  $1S$ & $6.437$ & I & $\bar{3}-3$ \\
  $1S$ & $6.450$ & II & $\bar{3}-3$\\
  $1S$ & $6.383$ & I & $6-\bar{6}$ \\
  $1S$ & $6.400$ & II & $6-\bar{6}$\\
  $1S$ & $6.276$ & Bag & $\bar{3}-3$ \\
  $1S$ & $6.252$ & Bag & $6-\bar{6}$ \\
\hline
  $1P$ & $6.718$ & I & $\bar{3}-3$ \\
  $1P$ & $6.714$ & II & $\bar{3}-3$\\
  $1P$ & $6.832$ & I & $6-\bar{6}$ \\
  $1P$ & $6.822$ & II & $6-\bar{6}$\\
\hline\hline
\end{tabular}
\end{center}
\end{table}

\begin{table}[H]
\caption{\label{tab:T4cCompNarrowSpin} Results for the spin shifts of
the $T_{4c}$ from Ref. \cite{narrow}.}
\begin{center}
\footnotesize\rm
\centering
\begin{tabular}{c c l r c c c}
\hline\hline
$N\ell$ & $M^{(0)}_{T4c}$ [GeV] & $J^{P(C)}$ & SS [GeV]  & LS + T [GeV] &
Model & Color \\
\hline
  $1S$ & $6.383$ & $0^+$ & $0.017$ & - & I & $6-\bar{6}$ \\
  $1S$ & $6.437$ & $0^+$ & $-0.011$ & - & I & $\bar{3}-3$ \\
  $1S$ & $6.437$ & $1^+$ & $0.003$ & - & I & $\bar{3}-3$ \\
  $1S$ & $6.437$ & $2^+$ & $0.032$ & - & I & $\bar{3}-3$ \\
\hline
  $1P$ & $6.832$ & $1^{--}$ & $0.011$ & 0 & I & $6-\bar{6}$ \\
  $1P$ & $6.718$ & $0^{-+}$ & $0.010$ & $-0.023$ & I & $\bar{3}-3$ \\
  $1P$ & $6.718$ & $1^{--}$ & $0.020$ & $-0.024$ & I & $\bar{3}-3$ \\
\hline\hline
\end{tabular}
\end{center}
\end{table}

First, we show the results of
Ref. \cite{narrow} in Tables \ref{tab:T4cCompNarrow} and
\ref{tab:T4cCompNarrowSpin}.
In this work a variational method with Gaussian trial wavefunctions was
employed to study all-heavy tetraquarks, using a four-body coordinate
system. The interactions were described with a potential due to the exchange
of color octets in two-body forces. Two potentials were used:
model I is a Cornell-type (Coulomb plus linear) and the model II is of
the form $A+Br^\beta$. Also a version of the MIT Bag model was used with
the Born-Oppenheimer approximation. Both color structures were considered,
$\bar{3}-3$ and $6-\bar{6}$. $S$-wave and $P$-wave were considered with both
potentials, and spin shifts were calculated with the Cornell-like potential.

In Table \ref{tab:T4cCompRussos} we compile the results of Refs.
\cite{mais1,mais2}, where the $T_{4c}$ production was studied. The estimates
for the $T_{4c}$ are very similar to those presented in this work.
The authors  used the diquark results of Ref. \cite{russos-BARION}, where
the $cc$ diquark was calculated as a baryon constituent (we also compared
these diquark results with ours). The same strategy
of dividing the problem into two-body problems was used, but only $S$-wave states
were calculated, and the spin-spin splitting was considered between each
spin 1/2 constituent pair, using the wavefunction at the origin of the
diquark or of the charmonium, depending on the  interacting pair. It is
interesting to see that the $0^{++}$ state is very close to our result, and
the $1^{+-}$ is also below the $\eta_c\,J/\psi$ threshold. However, the $2^{++}$
is about 20 MeV above the $J/\psi\,J/\psi$ threshold, indicating that this state
could be seen in the $J/\psi\,J/\psi$ invariant mass.
\begin{table}[H]
\caption{\label{tab:T4cCompRussos} Results for the $T_{4c}$ from
Refs. \cite{mais1,mais2}.}
\begin{center}
\footnotesize\rm
\centering
\begin{tabular}{c c c c l c c}
\hline\hline
$NL_T$ & $M^{(0)}_{T4c}$ [GeV] & $|\Psi(0)|$ [GeV$^{3/2}$] & $\langle r
\rangle$ [fm] &  $J^{PC}$  & $M^{f}_{T4c}$ [GeV] & Color \\
\hline
  $1S$ & 6.12 & 0.47 & 0.29 & $0^{++}$ & 5.97 & $\bar{3}-3$ \\
  $1S$ & 6.12 & 0.47 & 0.29 & $1^{+-}$ & 6.05 & $\bar{3}-3$ \\
  $1S$ & 6.12 & 0.47 & 0.29 & $2^{++}$ & 6.22 & $\bar{3}-3$ \\
  \hline \hline
\end{tabular}
\end{center}
\end{table}

In Table \ref{tab:T4cCompSwave} we  compare our results for the
$S$-wave $T_{4c}$ with those of the recent diquark-antidiquark studies:
those with antitriplet diquarks \cite{mais1,mais2}, and those with
the color-magnetic model \cite{WuT4c} and with QCD Sum Rules \cite{ChenT4c}.
\begin{table}[H]
\caption{\label{tab:T4cCompSwave} Comparison of our results for
the $S$-wave $T_{4c}$.}
\begin{center}
\footnotesize\rm
\centering
\begin{tabular}{c c c c c}
\hline\hline
 $J^{PC}$ & $M^{final}$ [GeV] & Ref. \cite{mais1,mais2} & Ref. \cite{WuT4c} &
Ref. \cite{ChenT4c} \\
\hline
$0^{++}$ & $5.9694$ & $5.966$ & $5.617-6.254$ & $6.44-7.15$ \\
$1^{+-}$ & $6.0209$ & $6.051$ & $5.720-6.137$ & $6.37-6.51$ \\
$2^{++}$ & $6.1154$ & $6.223$ & $5.777-6.194$ & $6.51-6.37$ \\
\hline \hline
\end{tabular}
\end{center}
\end{table}

In Table \ref{tab:T4cCompRosner} we compare our results with the contribution
of each term used to calculate the $0^{++}$ $T_{4c}$ in Ref.
\cite{Karliner}, which was based in meson and baryon masses.
The constant $V_0$ is obtained as twice the constant term $S$ obtained
from the fit of baryon and meson masses (which is added only into baryon
masses, related to the QCD string junction, as discussed in that reference).
Remember that in our model the spin-spin interaction is contained in the energy
eigenvalue.
\begin{table}[H]
\caption{\label{tab:T4cCompRosner} Comparison of our results for
the $0^{++}$ $T_{4c}$ with  Ref. \cite{Karliner}.}
\begin{center}
\footnotesize\rm
\centering
\begin{tabular}{c c c c c c c c}
\hline\hline
 $J^{PC}$ & $m_c$ [MeV] & $m_{cc}$ [MeV] & $E$ [MeV] & $V_0$ & $SS$ [MeV] & $M_{T4c}^{f}$ [MeV] &  \\
\hline
$0^{++}$ & $1462.2$ & $3133.4$ & $-297.3$ & $-$ & $(-106.0)$ & $5969.4$ & This work \\
$0^{++}$ & $1655.6$ & $3204.1$ & $-388.3$ & $330.2$ & $-158.5$ & $6191.5 \pm 25$ & Ref. \cite{Karliner} \\
\hline \hline
\end{tabular}
\end{center}
\end{table}

Finally, in Table \ref{tab:T4cCompPwave} we compare our results for the
$P$-wave $T_{4c}$ with  the old  diquark-antidiquark  predictions of
Chao \cite{Chao}, the recent diquark-antidiquark predictions of QCD Sum Rules
\cite{ChenT4c}  and with  lattice results  \cite{Chiu-Lattice}.
\begin{table}[H]
\caption{\label{tab:T4cCompPwave} Comparison of our results for
the $P$-wave $T_{4c}$.}
\begin{center}
\footnotesize\rm
\centering
\begin{tabular}{c c c c | c c}
\hline\hline
 $J^{PC}$ & $N^{2S_T+1}{L_T}_{J_T}$ & $M^{final}$ [GeV] & Ref. \cite{Chao} &
Ref. \cite{ChenT4c} & Ref. \cite{Chiu-Lattice} \\ \hline
$1^{--}$ & $1^1P_1$ & $6.5771$ & $6.55-6.82$ & $6.83-6.84$ & 6.420 \\
$1^{--}$ & $1^5P_1$ & $6.4954$ & $6.39$ &  & \\
\hline \hline
\end{tabular}
\end{center}
\end{table}

The use of the Cornell  potential allows us to study the charmonium spectrum
without the confining interactions, which can easily be ``switched off'' by
choosing the string tension to be zero. We can thus repeat all our calculations
and check whether we find bound diquark states and also a bound $T_{4c}$. We have
done these calculations and we find both diquark and tetraquark bound states.
The obtained diquark and $T_{4c}$ ground states have masses equal to
$m_{cc}=2881.4$ MeV and $T_{4c}=5.3-5.4$ GeV (for the lowest $1S$ states),
respectively, as shown in the Appendix.
These results can have applications in the context of relativistic heavy ion
collisions, where a deconfined medium is formed (the quark-gluon plasma, QGP).
Our results suggest that the $T_{4c}$ can be formed and perhaps survive in the QGP
phase.


\subsection{The role of $ 6 \, - \, \bar{6}$ configurations}

The tetraquark composed of four quarks of the same flavor is constrained by  the Pauli exclusion principle,
which restricts the possibilities of the diquark wave function. The most favorable case is the one presented in the previous
sections  where quarks in the diquark are in the attractive antitriplet color state (antisymmetric), in the ground state
$1S$ with no orbital nor radial excitations (symmetric) and with total spin $S=1$ (symmetric), resulting in an antisymmetric
wave function  appropriate for identical fermions. A diquark in the repulsive  color sextet configuration (symmetric)
should either have total spin $S=0$ (antisymmetric) or have an internal orbital excitation. This excitation strongly disfavors
the compactness of the diquarks, which  underlies the  assumption that the dynamics is dominated by one-gluon exchange.
Therefore, any internal orbital excitation in the diquarks can be safely neglected, and we end up with two orthogonal building
blocks: the antitriplet diquark with spin 1 and the sextet diquark  with spin 0.  There could be some mixing between these two
states. We know that spin 0 diquarks can only form
tetraquarks with total spin $S_T=0$, therefore the
tetraquarks composed of sextet diquarks would only mix with four of the 20 states
presented in this work, i. e. the states $1S$ and $1P$ with quantum numbers $J^{PC}=0^{++},\, 1^{--}$ and both respective
radial excitations.
All the other states are necessarily composed of pure antitriplet diquarks, since to have spin 1 sextet diquarks one would need
both diquark and antidiquark with one unit of internal orbital excitation, which is highly unlikely.

The exchange of one-gluon between a quark inside the diquark and an antiquark inside
the antidiquark could mix the color states $\bar{3}-3$ and $6-\bar{6}$.  Unfortunately this cannot be implemented in the
present model, where the four-body problem is factorized in subsequent two-body systems. The mixing can be taken  into account
in a full four-body problem, as done in Ref. \cite{SuHongLee}, where both color configurations
are present in the wave function from the beginning.
However, in this reference, as well as in most of the works in the literature, the sextet configuration is found to be negligible
when compared to the antitriplet configuration.  In the composite wave function the $\bar{3}-3$ component completely
dominates over the $6-\bar{6}$ one. This is essentially due to the repulsion inside the sextet diquark. The conclusion that we can draw
from this observation is that even though the $\bar{3}-3$ and $6-\bar{6}$ can mix,
a proper four-body approach reveals that they behave essentially as two independent states. The $6-\bar{6}$ contribution to the
$\bar{3}-3$ is expected to be negligible, and the former should be calculated separately as a pure $6-\bar{6}$ state.

Let us now present results for the pure $6 - \bar{6}$ tetraquark.
In our model, we  need first to compute the mass of the sextet diquark and then  calculate the tetraquark spectrum. The color
factor $\kappa_s$ of the Coulomb term in the potential corresponding to the sextet configurations is $+1/3$. The string tension $b$
of the linear confining term is the value taken for the  charmonium divided by four and its sign also changes. This interaction is
completely repulsive and clearly cannot yield a bound state. However one might argue that for non color-singlet configurations the long
distance part of the potential is not well known and might be confining.  We may get a rough estimate of  the sextet diquark mass as
being twice the charm quark mass.  The spin of the diquark is zero (and so is the spin of the tetraquark) and hence  all the spin-dependent
interactions  vanish. The interaction between a $6$ diquark and a $\bar{6}$ antidiquark is very attractive. The  color factor is $-10/3$,
and the  string tension is  positive and  a factor $10/4$ larger than  that of  charmonium. Using these parameters we can estimate the mass
of the $6-\bar{6}$ tetraquark in the four cases where it could mix with the $\bar{3}-3$ state. They are shown in Tables \ref{tab:66barWave} and
\ref{tab:66barMasses}.
We see that for the ground state we obtain
an extremely bound state around 4 GeV. This might be an  indication that the two-body approximation is already unrealistic and we should  take
into account the  finite size of the diquarks. If we use  a heavier diquark mass obtained with a confining string tension (about 3.2 GeV, 70 MeV
above the antitriplet diquark) we again find an extremely
bound  ground state, since increasing the diquark mass reduces the tetraquark size, increasing the contribution of
the attractive Coulomb term, (as happens when we move from charmonium to bottomonium).

%
%

\begin{table}[H]
\caption{\label{tab:66barWave} Results for $T_{4c}$ wavefunctions and
ground state ($1^1S_0$) diquark and antidiquark (sextet). Parameters are
$m_{cc} = 2m_c=2.9243$ MeV, $\alpha_s = 0.5202$, $b=10 \times b_{cc}/4=10 \times 0.1463/4$ GeV$^2$.}
\begin{center}
\footnotesize\rm
\centering
\begin{tabular}{c r r r r}
\hline\hline
    $N^{2S_T+1}L_T$ & $M^{(0)}$ [GeV] & $|R(0)|^2$ [GeV$^3$]& $\langle r^2 \rangle^{1/2}$ [fm] & $\displaystyle\left\langle \frac{v^2 }{c^2}\right\rangle$ \\
\hline
    $1^1S$ & $3.8611$ & $70.7780$ & $0.127$ & $0.820$ \\
\hline
    $1^1P$ & $5.8902$ & $0$ & $0.302$ & $0.341$ \\
\hline
    $2^1S$ & $6.0176$ & $16.3850$ & $0.368$ & $0.376$ \\
\hline
    $2^1P$ & $6.7567$ & $0$ & $0.539$ & $0.322$ \\
\hline\hline
\end{tabular}
\end{center}
\end{table}

\begin{table}[H]
\caption{\label{tab:66barMasses}   Results for $T_{4c}$ masses using
ground state ($1^1S_0$) diquarks (sextet - antisextet). Parameters are $m_{cc} = 2m_c=2.9243$ MeV,
$\alpha_s = 0.5202$, $b=10b_{cc}/4=10\times0.1463/4$ GeV$^2$.}
\begin{center}
\centering
\begin{tabular}{c r r r r r r r r r l}
\hline\hline
    $N^{2S_T+1}{L_T}_{J_T}$ & $\langle T \rangle$ & $\langle V^{(0)}_{V} \rangle$ & $\langle V^{(0)}_{S} \rangle$ & $\langle V^{(0)}_{SS} \rangle$ & $E^{(0)}$ & $M^{(0)}$ [MeV] & $\langle V^{(1)}_{LS} \rangle$ & $\langle V^{(1)}_T \rangle$ & $M^{f}$ [MeV] & $J^{PC}$ \\
\hline
    $1^1S_0$ &  $2397.0$ & $-4589.0$ & $204.6$ & $0$ & $-1987.6$ & $3861.1$ & $0$ & $0$ & $3861.1$ \\
\hline
    $1^1P_1$ & $996.1$ & $-1473.0$ & $518.9$ & $0$ & $41.6$ & $5890.2$ & $0$ & $0$ & $5890.2$ \\
\hline
    $2^1S_0$ & $1101.0$ & $-1566.0$ & $634.8$ & $0$ & $169.0$ & $6017.6$ & $0$ & $0$ & $6017.6$ \\
\hline
    $2^1P_1$ & $941.9$ & $-958.9$ & $925.0$ & $0$ & $908.1$ & $6756.7$ & $0$ & $0$ & $6756.7$ \\
\hline\hline
\end{tabular}
\end{center}
\end{table}

\

Before concluding  we would like to add a remark on the scale dependence of our results.
The one-loop QCD running coupling is given by:
$$
\alpha_s (Q^2) = \frac{ 12 \pi}{(33 - 2 N_f) \ln (Q^2 / \Lambda^2)}
$$
In this formula the scale $Q^2$ is an input. It is a choice which defines
the energy scale that is  relevant to the problem. In our case we have used
it as a constant, which was the same for the two-body $c \bar{c}$ problem and for
the $cc - \bar{c} \bar{c}$ two-body problem. We have here ignored the effects of the running  coupling.
In principle, we could have
chosen two different scales. For the $c \bar{c}$ problem it should be
$Q^2 \simeq m^2_c \simeq (1.4)^2$ GeV$^2$ and for the
$cc - \bar{c} \bar{c}$ problem it should be $Q^2 \simeq m^2_{cc}
\simeq (3.1)^2$  GeV$^2$. Using these numbers in the above formula we
obtain $\alpha_s(m_c) \simeq 0.5$ and $\alpha_s(m_{cc}) \simeq 0.35$.
Changing the scale, the running coupling is reduced by approximately
30 \%. Using $0.35$ instead of $0.5$ changes the resulting masses of the
bound states which come from the solution of the Schr\"odinger equation.
However, this change is only 5 \% for the lowest lying (1S) state. For
the higher states, i.e. the radial and orbital excitations, the effects
of the running coupling are even smaller, because the distance between the
two bodies is larger and the QCD-Coulomb interaction is less important. In
this region the uncertainties in the string tension are dominant. We have checked
that in the least favorable case (of a radial together with an orbital excitation)
for a diquark-antidiquark calculation, changing the string tension by
$\simeq 30$\%  leads to changes in the final $T_{4c}$ mass of $\simeq 3$\%.

An alternative way to  compute  running coupling effects was described
in Ref. \cite{eichten}, where the  authors compare (using their
formula 2.23)  $\alpha_s$ in bottomonium with $\alpha_s$ in charmonium with a
simple formula. Adapting their formula to our context, it reads:
$$
\alpha_s (T_{4c}) = \frac{ \alpha_s(\psi)}
{1 + \left[\displaystyle\frac{\alpha_s(\psi) }{12 \pi}\right] \left(33 - 2 N_f\right)
\ln\left(m^2_{T_{4c}}/m^2_{\psi}\right)}
$$
which then leads to the same results quoted above. The observation made above suggests
that we should correct our tables, changing the masses. Moreover, to be more accurate we
should also take into account the uncertainty in the scale choice, e.g., considering
$ Q^2 = \displaystyle\frac{m_c}{2} , m_c , 2 m_c$ and similarly for the diquark-antidiquark case.
However, we feel that this analysis would also imply a global uncertainty analysis,
which is beyond the scope of the present work. When dealing with very precise theoretical predictions,
all results should contain the theoretical errors, which reflect the uncertainties
in the calculations. This could be done in the present work by studying the
effects caused by changing the masses, couplings and string tension.
The uncertainty analysis could be improved by also including relativistic corrections
(a cubic term in the kinetic energy). However, this degree of precision would be more appropriate
when experimental data is available, allowing further constraints in the model, and it should be
postponed for future work.

This was the first calculation of the $T_{4c}$ spectrum with
a non-relativistic diquark-antidiquark model. It was meant to check
whether this approach reproduces what we know from the lattice
calculations, from QCD sum rules and from the results of the
Bethe-Salpeter approach. In this sense it is a preliminary calculation
which we believe has passed the test.
Further improvement could be made in the future by including a systematic analysis of the uncertainties, moving towards ``precision physics''.
The real novelty of this work would be the power to identify the components of the masses and determine the
role of the spin interactions, which are very difficult to isolate in
the lattice and in QCD sum rules calculations.

\section{Conclusion}
\label{Conclusion}
  \indent
In this work we first updated the Cornell model, (a very well known and
accepted model for charmonium) obtaining a
satisfactory reproduction of the charmonium spectrum, including the most
recently  measured states.   We then extended this model to  study
the all-charm tetraquark ($c \bar{c} c \bar{c}$).

We explored a diquark-antidiquark configuration, including $P$-wave tetraquarks,
and we extended the spin-dependent interactions between diquarks,
including a consistent strategy to deal with the tensor interaction between two
objects of spin 1. The fact that our model is relatively simple
compared to the four-body approach  and to the  relativistic models
allows us to study many of the $T_{4c}$ properties with clarity, especially the
role of the spin interactions.

We were able to study the behavior of the all-charm tetraquark when radial and orbital
excitations are included, investigating the contribution of the one-gluon
exchange, the confinement and the spin-dependent interactions, providing for
the first time detailed results which elucidate the dynamics of the
diquark-antidiquark structure.

The inclusion of one orbital excitation in the tetraquark also leads to a
significant increase in the possibilities of quantum numbers and to the
prediction of the exotic state with $J^{PC}=1^{-+}$. The orbitally-excited
$c\bar{c}$ states, having specific masses and quantum numbers, have different
decay channels, which may be investigated experimentally.

Our model is  simple and instructive, especially in what concerns spin interactions.
For the lowest $T_{4c}$ states our predictions are compatible with those made with
other approaches. For the higher states, in particular those with orbital
excitations, we make novel predictions which can be tested. In this region our
predictions are more reliable, since the diquark-antidiquark spatial separation
is bigger.

Our results and the others  found in the recent literature on
the $T_{4c}$ tetraquark, taken together, should encourage
a careful  experimental search for these states at LHCb and Belle II.


\section*{Acknowledgments}
\label{Acknowledgments}
  \indent

The authors acknowledge the support received from the brazilian funding agencies
FAPESP (contract 12/50984-4), CNPq and CAPES.
V. R. Debastiani also acknowledges the support from
Generalitat Valenciana in the Program Santiago Grisolia (Exp. GRISOLIA/2015/005).

\section*{Appendix}

In Table \ref{tab:T4c-OGE} we show the results for the lowest $1S$ tetraquark
states where both diquark and tetraquark were calculated without the linear
confinement term and hence the only interaction is one-gluon exchange.

These $1S$ states are deeply bound. As we can see in Table \ref{tab:T4c-OGE},
the binding energy for the lowest state, with $J^{PC} = 0^{++}$, is larger than
$-400$ MeV, where about 100 MeV come from the spin-spin interaction. The resulting
mass of $5.3$ GeV is compatible with the results of Ref. \cite{heupel}, where a
state with a dominant $\eta_c\eta_c$ component and mass $5.3 \pm (0.5)$ GeV
was found.

For the excited tetraquark states $2S$, $1P$ and $2P$, the binding energy is
around $-90$ MeV and the spin-dependent interactions are of the order of 10 MeV,
with masses around $5.6-5.7$ GeV.

\begin{table}[H]
\caption{\label{tab:T4c-OGE}  Results for lowest $T_{4c}$ states ($1S$) using
ground state ($1^3S_1$) diquarks. Both diquark and tetraquark are calculated
without the linear confinement term. Parameters are $m_{cc} = 2881.4$ MeV,
$\alpha_s = 0.5202$, $\mathbf{b=0}$, $\sigma=1.0831$ GeV.}
\begin{center}
\centering
\begin{tabular}{c r r r r r l r r}
\hline\hline
    $N^{2S_T+1}{L_T}_{J_T}$ & $\langle T \rangle$ & $\langle V^{(0)}_{V} \rangle$ & $\langle V^{(0)}_{SS} \rangle$ & $E^{(0)}$ & $M^{f}$ [MeV] & $J^{PC}$ & $\langle r^2 \rangle^{1/2}$ [fm] & $\displaystyle\left\langle \frac{v^2 }{c^2}\right\rangle$ \\
\hline
\hline
   \multicolumn{9}{c}{\textbf{Diquark}}\\
\hline
$1^3S_1$ & $41.4$ & $-85.3$ & $1.0$ & $-42.9$ & $2881.4$ & $1^+$ & $1.378$ & $0.028$ \\
\hline
   \multicolumn{9}{c}{\textbf{Tetraquark}}\\
\hline
$1^1S_0$ & $472.7$ & $-809.0$ & $-98.0$ & $-434.3$ & $5328.4$ &  $0^{++}$ & $0.291$ & $0.164$ \\
$1^3S_1$ & $408.0$ & $-751.9$ & $-44.0$ & $-387.8$ & $5374.9$ & $1^{+-}$ & $0.315$ & $0.142$ \\
$1^5S_2$ & $289.4$ & $-633.3$ & $33.4$ & $-310.4$ & $5452.3$ & $2^{++}$ & $0.374$ & $0.100$ \\
\hline\hline
\end{tabular}
\end{center}
\end{table}

\end{document}